\documentclass[twocolumn,twocolappendix]{aastex63}

\newcommand{\mc}{\mathcal{M}}

\received{\today}
\revised{}
\accepted{}
\submitjournal{ApJ}

\shorttitle{limits on smbhbs within 500\,mpc}
\shortauthors{The NANOGrav Collaboration}
\graphicspath{{./}{figures/}}

\defcitealias{Mingarelli2017}{M17}
\defcitealias{Schutz2016}{SM16}
\usepackage{graphicx}
\usepackage{subfigure}
\begin{document}

\title{The NANOGrav 11\,yr Data Set: Limits on Supermassive Black Hole Binaries in Galaxies within 500\,Mpc}

\correspondingauthor{Maria Charisi}
\email{mcharisi@caltech.edu}

\author{Zaven Arzoumanian}
\affiliation{X-Ray Astrophysics Laboratory, NASA Goddard Space Flight Center, Code 662, Greenbelt, MD 20771, USA}
\author[0000-0003-2745-753X]{Paul T. Baker}
\affiliation{Department of Physics and Astronomy, Widener University, One University Place, Chester, PA 19013, USA}
\author{Adam Brazier}
\affiliation{Cornell Center for Astrophysics and Planetary Science and Department of Astronomy, Cornell University, Ithaca, NY 14853, USA}
\author[0000-0003-3053-6538]{Paul R. Brook}
\affiliation{Department of Physics and Astronomy, West Virginia University, P.O. Box 6315, Morgantown, WV 26506, USA}
\affiliation{Center for Gravitational Waves and Cosmology, West Virginia University, Chestnut Ridge Research Building, Morgantown, WV 26505, USA}
\author[0000-0003-4052-7838]{Sarah Burke-Spolaor}
\affiliation{Department of Physics and Astronomy, West Virginia University, P.O. Box 6315, Morgantown, WV 26506, USA}
\affiliation{Center for Gravitational Waves and Cosmology, West Virginia University, Chestnut Ridge Research Building, Morgantown, WV 26505, USA}
\author{Bence Becsy}
\affiliation{Department of Physics, Montana State University, Bozeman, MT 59717, USA}
\author[0000-0003-3579-2522]{Maria Charisi$^{\color{magenta}\S}$}
\altaffiliation{NANOGrav Physics Frontiers Center Postdoctoral Fellow}
\affiliation{Theoretical AstroPhysics Including Relativity (TAPIR), MC 350-17, California Institute of Technology, Pasadena, California 91125, USA}
\affiliation{Department of Physics and Astronomy, Vanderbilt University, 2301 Vanderbilt Place, Nashville, TN 37235, USA}
\author[0000-0002-2878-1502]{Shami Chatterjee}
\affiliation{Cornell Center for Astrophysics and Planetary Science and Department of Astronomy, Cornell University, Ithaca, NY 14853, USA}
\author[0000-0002-4049-1882]{James M. Cordes}
\affiliation{Cornell Center for Astrophysics and Planetary Science and Department of Astronomy, Cornell University, Ithaca, NY 14853, USA}
\author[0000-0002-7435-0869]{Neil J. Cornish}
\affiliation{Department of Physics, Montana State University, Bozeman, MT 59717, USA}
\author[0000-0002-2578-0360]{Fronefield Crawford}
\affiliation{Department of Physics and Astronomy, Franklin \& Marshall College, P.O. Box 3003, Lancaster, PA 17604, USA}
\author[0000-0002-6039-692X]{H. Thankful Cromartie}
\affiliation{University of Virginia, Department of Astronomy, P.O. Box 400325, Charlottesville, VA 22904, USA}
\author[0000-0002-2185-1790]{Megan E. DeCesar}
\affiliation{Department of Physics, Lafayette College, Easton, PA 18042, USA}
\author[0000-0002-6664-965X]{Paul B. Demorest}
\affiliation{National Radio Astronomy Observatory, 1003 Lopezville Rd., Socorro, NM 87801, USA}
\author[0000-0001-8885-6388]{Timothy Dolch}
\affiliation{Department of Physics, Hillsdale College, 33 E. College Street, Hillsdale, Michigan 49242, USA}
\author{Rodney D. Elliott}
\affiliation{Department of Astrophysical and Planetary Sciences, University of Colorado Boulder, Boulder, CO 80309, USA}
\author{Justin A. Ellis}
\affiliation{Infinia ML, 202 Rigsbee Avenue, Durham NC, 27701}
\author{Elizabeth C. Ferrara}
\affiliation{NASA Goddard Space Flight Center, Greenbelt, MD 20771, USA}
\author[0000-0001-8384-5049]{Emmanuel Fonseca}
\affiliation{Department of Physics, McGill University, 3600  University St., Montreal, QC H3A 2T8, Canada}
\author[0000-0001-6166-9646]{Nathan Garver-Daniels}
\affiliation{Department of Physics and Astronomy, West Virginia University, P.O. Box 6315, Morgantown, WV 26506, USA}
\affiliation{Center for Gravitational Waves and Cosmology, West Virginia University, Chestnut Ridge Research Building, Morgantown, WV 26505, USA}
\author[0000-0001-8158-638X]{Peter A. Gentile}
\affiliation{Department of Physics and Astronomy, West Virginia University, P.O. Box 6315, Morgantown, WV 26506, USA}
\affiliation{Center for Gravitational Waves and Cosmology, West Virginia University, Chestnut Ridge Research Building, Morgantown, WV 26505, USA}
\author[0000-0003-1884-348X]{Deborah C. Good}
\affiliation{Department of Physics and Astronomy, University of British Columbia, 6224 Agricultural Road, Vancouver, BC V6T 1Z1, Canada}
\author[0000-0003-2742-3321]{Jeffrey S. Hazboun}
\affiliation{Physical Sciences Division, University of Washington Bothell, 18115 Campus Way NE, Bothell, WA 98011, USA}
\author{Kristina Islo}
\affiliation{Center for Gravitation, Cosmology and Astrophysics, Department of Physics, University of Wisconsin-Milwaukee,\\ P.O. Box 413, Milwaukee, WI 53201, USA}
\author[0000-0003-1082-2342]{Ross J. Jennings}
\affiliation{Cornell Center for Astrophysics and Planetary Science and Department of Astronomy, Cornell University, Ithaca, NY 14853, USA}
\author[0000-0001-6607-3710]{Megan L. Jones}
\affiliation{Center for Gravitation, Cosmology and Astrophysics, Department of Physics, University of Wisconsin-Milwaukee,\\ P.O. Box 413, Milwaukee, WI 53201, USA}
\author[0000-0002-3654-980X]{Andrew R. Kaiser}
\affiliation{Department of Physics and Astronomy, West Virginia University, P.O. Box 6315, Morgantown, WV 26506, USA}
\affiliation{Center for Gravitational Waves and Cosmology, West Virginia University, Chestnut Ridge Research Building, Morgantown, WV 26505, USA}
\author[0000-0001-6295-2881]{David L. Kaplan}
\affiliation{Center for Gravitation, Cosmology and Astrophysics, Department of Physics, University of Wisconsin-Milwaukee,\\ P.O. Box 413, Milwaukee, WI 53201, USA}
\author[0000-0002-6625-6450]{Luke Zoltan Kelley}
\affiliation{Center for Interdisciplinary Exploration and Research in Astrophysics (CIERA), Northwestern University, Evanston, IL 60208}
\author[0000-0003-0123-7600]{Joey Shapiro Key}
\affiliation{Physical Sciences Division, University of Washington Bothell, 18115 Campus Way NE, Bothell, WA 98011, USA}
\author[0000-0003-0721-651X]{Michael T. Lam}
\affiliation{School of Physics and Astronomy, Rochester Institute of Technology, Rochester, NY 14623, USA}
\affiliation{Laboratory for Multiwavelength Astrophysics, Rochester Institute of Technology, Rochester, NY 14623, USA}
\author{T. Joseph W. Lazio}
\affiliation{Jet Propulsion Laboratory, California Institute of Technology, 4800 Oak Grove Drive, Pasadena, CA 91109, USA}
\affiliation{Theoretical AstroPhysics Including Relativity (TAPIR), MC 350-17, California Institute of Technology, Pasadena, California 91125, USA}
\author{Jing Luo}
\affiliation{Department of Astronomy \& Astrophysics, University of Toronto, 50 Saint George Street, Toronto, ON M5S 3H4, Canada}
\author[0000-0001-5229-7430]{Ryan S. Lynch}
\affiliation{Green Bank Observatory, P.O. Box 2, Green Bank, WV 24944, USA}
\author{Chung-Pei Ma}
\affiliation{Department of Astronomy, University of California Berkeley, Campbell Hall, Berkeley, CA 94720}
\author[0000-0003-2285-0404]{Dustin R. Madison}
\altaffiliation{NANOGrav Physics Frontiers Center Postdoctoral Fellow}
\affiliation{Department of Physics and Astronomy, West Virginia University, P.O. Box 6315, Morgantown, WV 26506, USA}
\affiliation{Center for Gravitational Waves and Cosmology, West Virginia University, Chestnut Ridge Research Building, Morgantown, WV 26505, USA}
\author[0000-0001-7697-7422]{Maura A. McLaughlin}
\affiliation{Department of Physics and Astronomy, West Virginia University, P.O. Box 6315, Morgantown, WV 26506, USA}
\affiliation{Center for Gravitational Waves and Cosmology, West Virginia University, Chestnut Ridge Research Building, Morgantown, WV 26505, USA}
\author[0000-0002-4307-1322]{Chiara M. F. Mingarelli}
\affiliation{Center for Computational Astrophysics, Flatiron Institute, 162 5th Avenue, New York, New York, 10010, USA}
\affiliation{Department of Physics, University of Connecticut, 196 Auditorium Road, U-3046, Storrs, CT 06269-3046, USA}
\author[0000-0002-3616-5160]{Cherry Ng}
\affiliation{Dunlap Institute for Astronomy and Astrophysics, University of Toronto, 50 St. George St., Toronto, ON M5S 3H4, Canada}
\author[0000-0002-6709-2566]{David J. Nice}
\affiliation{Department of Physics, Lafayette College, Easton, PA 18042, USA}
\author[0000-0001-5465-2889]{Timothy T. Pennucci}
\altaffiliation{NANOGrav Physics Frontiers Center Postdoctoral Fellow}
\affiliation{National Radio Astronomy Observatory, 520 Edgemont Road, Charlottesville, VA 22903, USA}
\affiliation{Institute of Physics, E\"{o}tv\"{o}s Lor\'{a}nd University, P\'{a}zm\'{a}ny P. s. 1/A, 1117 Budapest, Hungary}
\author[0000-0002-8826-1285]{Nihan S. Pol}
\affiliation{Department of Physics and Astronomy, West Virginia University, P.O. Box 6315, Morgantown, WV 26506, USA}
\affiliation{Department of Physics and Astronomy, Vanderbilt University, 2301 Vanderbilt Place, Nashville, TN 37235, USA}
\affiliation{Center for Gravitational Waves and Cosmology, West Virginia University, Chestnut Ridge Research Building, Morgantown, WV 26505, USA}
\author[0000-0001-5799-9714]{Scott M. Ransom}
\affiliation{National Radio Astronomy Observatory, 520 Edgemont Road, Charlottesville, VA 22903, USA}
\author[0000-0002-5297-5278]{Paul S. Ray}
\affiliation{Space Science Division, Naval Research Laboratory, Washington, DC 20375-5352, USA}
\author[0000-0002-7283-1124]{Brent J. Shapiro-Albert}
\affiliation{Department of Physics and Astronomy, West Virginia University, P.O. Box 6315, Morgantown, WV 26506, USA}
\affiliation{Center for Gravitational Waves and Cosmology, West Virginia University, Chestnut Ridge Research Building, Morgantown, WV 26505, USA}
\author[0000-0002-7778-2990]{Xavier Siemens}
\affiliation{Department of Physics, Oregon State University, Corvallis, OR 97331, USA}
\affiliation{Center for Gravitation, Cosmology and Astrophysics, Department of Physics, University of Wisconsin-Milwaukee,\\ P.O. Box 413, Milwaukee, WI 53201, USA}
\author[0000-0003-1407-6607]{Joseph Simon}
\affiliation{Jet Propulsion Laboratory, California Institute of Technology, 4800 Oak Grove Drive, Pasadena, CA 91109, USA}
\affiliation{Theoretical AstroPhysics Including Relativity (TAPIR), MC 350-17, California Institute of Technology, Pasadena, California 91125, USA}
\author[0000-0002-6730-3298]{Ren\'{e}e Spiewak}
\affiliation{Centre for Astrophysics and Supercomputing, Swinburne University of Technology, P.O. Box 218, Hawthorn, Victoria 3122, Australia}
\author[0000-0001-9784-8670]{Ingrid H. Stairs}
\affiliation{Department of Physics and Astronomy, University of British Columbia, 6224 Agricultural Road, Vancouver, BC V6T 1Z1, Canada}
\author[0000-0002-1797-3277]{Daniel R. Stinebring}
\affiliation{Department of Physics and Astronomy, Oberlin College, Oberlin, OH 44074, USA}
\author[0000-0002-7261-594X]{Kevin Stovall}
\affiliation{National Radio Astronomy Observatory, 1003 Lopezville Rd., Socorro, NM 87801, USA}
\author[0000-0002-1075-3837]{Joseph K. Swiggum}
\altaffiliation{NANOGrav Physics Frontiers Center Postdoctoral Fellow}
\affiliation{Department of Physics, Lafayette College, Easton, PA 18042, USA}
\author[0000-0003-0264-1453]{Stephen R. Taylor}
\affiliation{Department of Physics and Astronomy, Vanderbilt University, 2301 Vanderbilt Place, Nashville, TN 37235, USA}
\author[0000-0002-4162-0033]{Michele Vallisneri}
\affiliation{Jet Propulsion Laboratory, California Institute of Technology, 4800 Oak Grove Drive, Pasadena, CA 91109, USA}
\affiliation{Theoretical AstroPhysics Including Relativity (TAPIR), MC 350-17, California Institute of Technology, Pasadena, California 91125, USA}
\author[0000-0003-4700-9072]{Sarah J. Vigeland}
\affiliation{Center for Gravitation, Cosmology and Astrophysics, Department of Physics, University of Wisconsin-Milwaukee,\\ P.O. Box 413, Milwaukee, WI 53201, USA}

\author[0000-0002-6020-9274]{Caitlin A. Witt}
\affiliation{Department of Physics and Astronomy, West Virginia University, P.O. Box 6315, Morgantown, WV 26506, USA}
\affiliation{Center for Gravitational Waves and Cosmology, West Virginia University, Chestnut Ridge Research Building, Morgantown, WV 26505, USA}

\collaboration{1000}{The NANOGrav Collaboration}
\noaffiliation



\begin{abstract}

Supermassive black hole binaries (SMBHBs) should form frequently in galactic nuclei as a result of galaxy mergers. At sub-parsec separations, binaries become strong sources of low-frequency gravitational waves (GWs), targeted by Pulsar Timing Arrays (PTAs). We used recent upper limits on continuous GWs from the North American Nanohertz Observatory for Gravitational Waves (NANOGrav) 11\,yr dataset to place constraints on putative SMBHBs in nearby massive galaxies. 
We compiled a comprehensive catalog of $\sim$44,000 galaxies in the local universe (up to redshift $\sim$0.05) and populated them with hypothetical binaries, assuming that the total mass of the binary is equal to the SMBH mass derived from global scaling relations. Assuming circular equal-mass binaries emitting at NANOGrav's most sensitive frequency of 8\,nHz, we found that 216 galaxies are within NANOGrav's sensitivity volume.
We ranked the potential SMBHBs based on GW detectability by calculating the total signal-to-noise ratio (S/N) such binaries would induce within the NANOGrav array.
We placed constraints on the chirp mass and mass ratio of the 216 hypothetical binaries. For 19 galaxies, only very unequal-mass binaries are allowed, with the mass of the secondary less than 10 percent that of the primary, roughly comparable to constraints on a SMBHB in the Milky Way.
Additionally, we were able to exclude binaries delivered by major mergers (mass ratio of at least 1/4) for several of these galaxies. We also derived the first limit on the density of binaries delivered by major mergers purely based on GW data.

\end{abstract}

\keywords{editorials, notices --- 
miscellaneous --- catalogs --- surveys}


\section{Introduction}
\label{section:Introduction}

Supermassive black hole binaries are a natural consequence of galaxy mergers, since massive galaxies are known to host supermassive black holes (SMBHs) \citep{Kormendy2013,Haehnelt2002}. In the post-merger galaxy, the SMBHs evolve through dynamical friction and stellar interactions, and 
eventually form a gravitationally bound binary.
At small (milli-parsec) separations, the binary evolution is dominated by the emission of gravitational waves (GWs). Even though the main steps of binary evolution were proposed decades ago \citep{Begelman1980}, there are significant theoretical uncertainties in the path of the SMBHs to their final coalescence (e.g., see \citealt{Colpi2014,DeRosa2019} for reviews). Therefore, detecting binaries in the GW regime is of paramount importance for understanding galaxy evolution \citep{2019BAAS...51c.336T}.

The most massive binaries ($10^8-10^{10}\,M_{\odot}$) are strong sources of GWs in the nanohertz band ($\sim 1-100$\,nHz) and are promising targets for Pulsar Timing Arrays (PTAs). PTAs measure precisely the times of arrival (TOAs) of pulses for a set of very stable millisecond pulsars. GWs passing through the Milky Way distort the Earth-pulsar distance, and induce coherent deviations in the TOAs. If such deviations are measured over multiple pulsars in the array with a characteristic quadrupole correlation signature, then GWs can be detected  \citep{Detweiler1979,Hellings1983}.

Currently, three collaborations are involved in the effort to detect nanohertz GWs: the North American Nanohertz Observatory for Gravitational waves 
\citep[\hbox{NANOGrav},][]{2019BAAS...51g.195R}, the European Pulsar Timing Array \citep[\hbox{EPTA},][]{Desvignes2016} and the Parkes Pulsar Timing Array \citep[\hbox{PPTA},][]{krh+20}. Together along with other emerging PTAs, they form a wider consortium, known as the International Pulsar Timing Array \citep[\hbox{IPTA},][]{pdd+19}.

PTAs are expected to detect two main types of GW signals: (1)~the stochastic GW background (GWB), i.e., the incoherent superposition of many unresolved signals from a population of  binaries, and (2)~continuous waves (CWs), i.e., monochromatic GWs from individual SMBHBs. Additional (albeit less likely) signals, such as bursts with memory from SMBHBs, primordial GWs, or cosmic strings may also be detected (\citealt{Burke-Spolaor2019}). 
As the sensitivity of PTAs is continuously improving, the first detection is expected in the near future \citep{Rosado2015,Taylor2016,Mingarelli2017,Kelley2018}. 

Even before the first detection, PTA upper limits on the GWB already provide interesting astrophysical insights. For instance, \citet{Sesana2018} demonstrated that the GWB inferred from the population of binary candidates identified as quasars with periodic variability \citep{Graham2015,Charisi2016} is in tension with the current upper limits. This suggests that the samples of photometrically identified candidates likely contain false detections. \citet{Holgado2018} reached a similar conclusion for the population of blazars with periodic variability \citep{Sandrinelli2018}.

Similarly, the upper limits on CWs provide constraints on binary candidates. For instance, a SMBHB 
was suggested in the galaxy \hbox{3C~66B}, because the radio core of the galaxy shows an elliptical motion \citep{Sudou2003} with a period of 1.05\,yr. Since this is a relatively nearby system (at $\sim$85\,Mpc),  it could produce GWs detectable by PTAs. 
Our recent study using the NANOGrav 11\,yr dataset constrained the chirp mass of the potential SMBHB to less
than $1.65\times10^9 M_{\odot}$ \citep{Multimessenger_NANOGrav}, improving by a factor of four on previous limits based on a single pulsar \citep{Jenet2004}.

Furthermore, \citet{Schutz2016} [hereafter \citetalias{Schutz2016}]
examined two samples of local galaxies for putative binaries. They focused on
galaxies with SMBHs, the mass of which has been measured dynamically \citep{McConnell2013}, and galaxies from the MASSIVE survey, which consists of the most massive early-type galaxies within 100\,Mpc \citep{Ma2014}.
Based on upper limits from the PPTA's first data release \citep{Zhu2014}, and the most recent data release from EPTA \citep{Babak2016}, they placed contraints on the mass ratio of hypothetical binaries in the above sample. They concluded that for a handful of nearby massive galaxies only very unequal-mass binaries are plausible at PPTA's and EPTA's sensitive frequencies, with the mass of the secondary SMBH a few percent that of the primary.

In this paper, we expand on this work considering the most recent (and most sensitive) CW upper limits from NANOGrav, derived from the 11\,yr dataset \citep{NANOGrav_CWs_2018,NANOGrav_11yr_2018}. This dataset spans from 2004 to 2015 and contains TOAs of 45 pulsars. Of those, only 34 were observed for at least 3\,yr and were included in the present CW analysis. At the most sensitive frequency ($f_{\mathrm{GW}} = 8$\,nHz), the sky-averaged upper limit on the GW strain is $h_0 < 7.3\times10^{-15}$.\footnote{From the same dataset, the limit on the amplitude of the GWB, at a frequency of $f=1 \mathrm{yr}^{-1}$ for a fiducial power-law GWB spectrum with a slope of -2/3  is $A_{\mathrm{GWB}}<1.45\times10^{-15}$ \citep{NANOGrav_GWB_2018}.} For comparison, the sky-averaged upper limits used in \citetalias{Schutz2016} were $h_0<1.7\times10^{-14}$ and $h_0<1.1\times10^{-14}$ (at 10\,nHz) from PPTA and EPTA, respectively. 

With the NANOGrav 11\,yr upper limits, we have already excluded the presence of a circular equal-mass binary, with chirp mass $\mathcal{M} > 1.6\times10^{9} M_{\odot}$, emitting in the frequency range between 2.8\,nHz  to 317.8\,nHz, in the Virgo Cluster \citep{NANOGrav_CWs_2018}. Here, we extend the astrophysical inference based on the 11\,yr dataset beyond the Virgo Cluster. For this, we construct a comprehensive galaxy catalog, which contains all the massive nearby galaxies, using the Two Micron All-Sky Survey (2MASS).  We carefully determine the galaxy distances and the masses of their SMBHs. We select a subset of 216 galaxies which are in the sensitivity volume of NANOGrav, i.e., they would emit detectable GWs if they hosted equal mass circular binaries at 8\,nHz. Note that the galaxies examined in \citetalias{Schutz2016} are a subset of catalog that we use here. 

We rank the above galaxies based on the GW detectability, calculating the total S/N the tentative SMBHBs induce to the NANOGrav pulsars. Subsequently, we run the upper limits analysis in the direction of the galaxies and convert the GW strain limits to constraints on the mass ratio (using the distance and the mean SMBH mass from the catalog we constructed). For the five top ranked galaxies, we derive upper limits sampling directly on the mass ratio, incorporating the uncertainty in the total mass. We also place constraints on binaries delivered by major mergers, examining frequencies for which binaries with a mass ratio of $q>1/4$ can be excluded and derive a limit on the number density of such binaries at 8\,nHz.

The paper is organized as follows. In \S~\ref{sec:catalog}, we describe the properties of the galaxy catalog and the selection of the sub-sample of galaxies that lie within the NANOGrav sensitivity volume.  
In \S~\ref{sec:GW_Analysis}, we summarize the statistical analysis of the NANOGrav data. In \S~\ref{sec:results}, we present the constraints on putative SMBHBs in these galaxies. We compare our results with previous related studies and discuss future directions in \S~\ref{sec:discuss} followed by a brief summary in \S~\ref{sec:summary}.

\section{Galaxy Catalog}\label{sec:catalog}

We compiled a comprehensive catalog of nearby galaxies. Our starting point was the 2MASS Redshift Survey (\hbox{2MRS}, \citealt{Huchra2012}), which consists of 43,533 galaxies, the redshifts of which have been measured spectroscopically.
Our choice of 2MASS for constructing a census of galaxies in the local Universe was motivated by the fact that, as a near-IR survey ($K$-band; 2.2$\mu m$), it is minimally affected by interstellar extinction. In fact, 2MRS is 97.6\% complete to a distance of 300\,Mpc, and magnitudes $m_K \leq 11.75$ \citep{Huchra2012}.

In order to determine whether NANOGrav can probe a tentative binary in the above galaxies, we needed to calculate the GW strain such a binary would produce. The amplitude of the GW strain $h_{\mathrm{GW}}$ is given by 
\begin{equation}
\label{eq:strain}
h_{\mathrm{GW}}=\frac{2 (G\mathcal{M})^{5/3} (\pi f_{\mathrm{GW}})^{2/3}}{c^4 D},
\end{equation}
$D$ is the luminosity distance, $f_{\mathrm{GW}}$ the observed GW frequency, and $\mathcal{M}$ the chirp mass,
\begin{equation}
\label{eq:chirp_mass}
    \mathcal{M}=\left[\frac{q}{(1+q)^2}\right]^{3/5} M_{\mathrm{tot}},
\end{equation}
where $M_{\mathrm{tot}}$ is the total mass and $q$ the mass ratio (e.g., see \citealt{Finn2000}). The observed frequency and chirp mass are related to the rest-frame values as $f_r=f_{\mathrm{GW}}\times(1+z)$ and $\mathcal{M}_{r}=\mathcal{M}/(1+z)$. However, since our catalog includes only relatively nearby galaxies ($z<0.05$), we set $1+z\simeq1$. We note that as the sensitivities of PTAs improve, and in turn the distances to which they can probe increase, redshifts should be incorporated in future analyses.

Therefore, for each galaxy an estimate of the distance and the mass of the SMBH (which we attributed to the total mass of the binary) was required. Since at milli-parsec separations, it is almost impossible to directly resolve two SMBHs \citep{2018ApJ...863..185D}, assigning the mass of a central SMBH (either measured or estimated from scaling relations) to be the total mass of the binary is the best possible alternative. 
Below we describe how we estimated these quantities.

\subsection{Galaxy Distances}
\label{sec:dist}

In the local volume, converting redshift to distance is not trivial because the spectroscopically measured velocities are significantly affected by the motion of galaxies (e.g., due to infall to clusters). Therefore, it is necessary to correct for the distortions in the local velocity field or use direct distance measurements which are available for several local galaxies. For the latter, we used the following catalogs:
(A)~Cosmicflows-3 \citep{Tully2016} extracted from the the Extragalactic Distance Database (\citealt{Tully2009}), and (B)~the galaxy groups catalog  \citep{Crook2007}. 

Cosmicflows-3 is a compilation of distances of nearby galaxies, which were obtained with high quality methods. These include methods that rely on standard candles,
such as: (1)~the Cepheid period-luminosity relation, in which the intrinsic luminosity is determined from the pulsation period of the star; (2)~the Tip of the Red Giant Branch, in which the brightest red giants in the Hertzsprung-Russel diagram are used as standard candles; and (3)~Type~Ia~supernovae. Cosmicflows-3 also incorporates distances from additional methods, including: (4)~the Tully-Fisher relation, an empirical correlation between the rotation velocity of spiral galaxies and their intrinsic luminosity; (5)~the fundamental plane of elliptical galaxies, a bivariate correlation between the luminosity, the effective radius and the dispersion velocity of the galaxy; and (6)~the surface brightness fluctuations method, which relies on the variance of the observed light distribution (e.g., distant galaxies have lower variance across pixels and appear smoother). 
Among the galaxies in the 2MRS sample, 8,625 were covered in Cosmicflows-3.

For galaxies not included in Cosmicflows-3, we used the distance estimates from the catalog of galaxy groups. In this catalog, a ``friends-of-friends" algorithm was employed to determine galaxy groups by identifying neighboring galaxies with similar distances. 
The galaxy distances were determined by applying a flow-field model, following \citet{Mould2000}. More specifically, the heliocentric velocities were first converted to the Local Group frame. Subsequently, the galaxies in the vicinity of major local clusters (i.e., the Virgo Cluster, the Shapley Supercluster, and the Great Attractor) were assigned the velocity of the respective cluster (see \citealt{Crook2007} and \citealt{Mould2000} for details). Finally, an additional correction was added to account for the gravitational pull towards the aforementioned major galaxy clusters.
We used the High-Density-Contrast (HDC) catalog from \citet{Crook2007}, in which a galaxy was considered a group member when the density contrast was 80 or more, and for groups with more than two members, we used the mean distance of the members.
The group catalog provided the distances for another 4,533 galaxies.

For the remaining galaxies (i.e., galaxies not included in any of the above two catalogs), we followed the prescription from \citet{Mould2000}, described above,
to correct for velocity distortions in the local volume. We adopted a nominal value of $H_0= 70\,\mathrm{km}\,\mathrm{s}^{-1}\,\mathrm{Mpc}^{-1}$, and accordingly adjusted the distances from Cosmicflows-3 and galaxy groups, in which the adopted values were 75\,km~s${}^{-1}$~Mpc${}^{-1}$ and 73\,km~s${}^{-1}$~Mpc${}^{-1}$, respectively. In our calculations, we did not include the distance uncertainties, primarily because they are negligible compared to the uncertainties of the mass estimates (see below).

\subsection{SMBH Masses}
\label{sec:BH_mass}

For the SMBH masses, we used direct measurements when available and global scaling relationships otherwise. Dynamical mass measurements are observationally expensive, and are limited only to a small number of galaxies. We used the catalog compiled by \citet{McConnell2013}, which includes 72 galaxies with high-quality dynamical mass measurements, from observations of stellar, gaseous, or maser kinematics. We added to this catalog recent measurements for the galaxies NGC~4526, M60-UCD1, NGC~1271, NGC~1277, and NGC~1600 \citep{Davis2013,Seth2014,Walsh2015,Walsh2016,Thomas2016}. We also included 29 galaxies from the Active Galactic Nuclei (AGN) Black Hole Mass Database,\footnote{\url{http://www.astro.gsu.edu/AGNmass/}} the masses of which were determined with spectroscopic reverberation mapping \citep{Bentz2015}.
However, the above dynamical tracers probe spatial scales that cannot distinguish if the mass corresponds to one or two SMBHs, and thus reflect the total enclosed mass.

For the remaining galaxies, we estimated the SMBH masses from correlations with observable quantities, like the $M_{\mathrm{BH}}$-$\sigma$ or the $M_{\mathrm{BH}}$-$M_{\mathrm{bulge}}$ correlations.
First, we extracted measurements of the velocity dispersion~$\sigma$ for 3,300 galaxies from the Lyon-Meudon Extragalactic Database (LEDA), using the Extragalactic Distance Database to cross-correlate our sample with \hbox{LEDA}. We used the $M_{\mathrm{BH}}$-$\sigma$ relation from \citet{McConnell2013}
\begin{equation}
    \log_{10}(M_{\mathrm{BH}})=a_1 + b_1 \log_{10}(\sigma/200\,\mathrm{km\,s}^{-1}) \,,
\end{equation}
with $(a_1,b_1)=(8.39,5.20)$ for early-type (elliptical and S0) galaxies and $(a_1,b_1)=(8.07,5.06)$ for late-type (spiral) galaxies.
Of the 3,300 galaxies, we excluded 810 galaxies because their velocities were beyond the range from which the correlation was established (80\,km~s${}^{-1}$--400\,km~s${}^{-1}$) or their dispersion measurements had significant uncertainty ($>$10\%). Additionally, the morphological types of 284 galaxies were not included in the catalog or they were classified as irregular. For this subset, we did not use the $M_{\mathrm{BH}}$-$\sigma$ relation.

If a measurement of the dispersion velocity was not available (or we excluded it for reasons mentioned above), we used the $M_{\mathrm{BH}}$-$M_{\mathrm{bulge}}$ relation to estimate the SMBH mass.
We first used the $K$-band magnitude to estimate the stellar mass, which in turn was used to calculate the bulge mass. In particular, we determined the absolute magnitude $M_K$ from the observed $m_K$ as
\begin{equation}
    M_K = m_{K}-5\times \log_{10}(D)-25-0.11\times A_v,
\end{equation}
where $D$ is the distance in Mpc and $A_v$ is the extragalactic extinction, which we calculated from the reddening relation $R_V=A_V/E_{\mathrm{B-V}}=3.1$ \citep{Fitzpatrick1999}, with the color term $E_{\mathrm{B-V}}$ provided by the 2MRS catalog.  Next, we obtained the total stellar mass of the galaxy $M_*$ using the correlation \citep{Cappellari2013} 
\begin{equation}
    \log_{10}(M_*)= 10.58-0.44\times(M_K +23) \,.
\end{equation}
At this step, we excluded 21 quasars, since the $K$-band luminosity does not correspond to the stellar luminosity, but is dominated by dust emission \citep{Barvainis1987}.

\begin{table}
\caption{Summary of methods for estimating SMBH masses. In the second part of the table, we focus on the subset of galaxies the SMBH mass of which was determined from the $M_{\mathrm{BH}}$-$M_{\mathrm{bulge}}$ correlation. We show the adopted mean bulge-to-total mass ratio ($f_{\mathrm bulge}$) as a function of galaxy type from \citet{Weinzirl2009}.}\label{Table:Galaxy_type}
\begin{tabular}{lll}
\hline
Method& Number of galaxies&\\
\hline
Dynamical methods&\hspace{0.5cm}77&\\ Reverberation mapping&\hspace{0.5cm}29&\\
Excluded quasars&\hspace{0.5cm}21&\\
$M_{\mathrm{BH}}$-$\sigma$&\hspace{0.5cm}2,206&\\
$M_{\mathrm{BH}}$-$M_{\mathrm{bulge}}$&\hspace{0.5cm}41,200&\\
\end{tabular}

\begin{tabular}{llll}
\hline \hline

&Galaxy Type&Number of galaxies&$f_{\mathrm bulge}$\\\hline
&Unknown type$^*$&\hspace{1cm}18,575&1 and 0.31\\
&Known type&\hspace{1cm}22,625\\\hline
&\hspace{0.5cm}Elliptical&\hspace{1cm}8,791 &\hspace{0.5cm}1.00\\
&\hspace{0.5cm}S0&\hspace{1cm}1,363 &\hspace{0.5cm}0.25\\
&\hspace{0.5cm}Sa&\hspace{1cm}2,237&\hspace{0.5cm}0.31\\
&\hspace{0.5cm}Sab&\hspace{1cm}1,626&\hspace{0.5cm}0.29\\
&\hspace{0.5cm}Sb&\hspace{1cm}2,709&\hspace{0.5cm}0.17\\
&\hspace{0.5cm}Sbc&\hspace{1cm}1,908&\hspace{0.5cm}0.13\\
&\hspace{0.5cm}Sc&\hspace{1cm}2,838&\hspace{0.5cm}0.17\\
&\hspace{0.5cm}Scd&\hspace{1cm}693&\hspace{0.5cm}0.09\\
&\hspace{0.5cm}Sd&\hspace{1cm}221&\hspace{0.5cm}0.14\\
&\hspace{0.5cm}Sdm&\hspace{1cm}136&\hspace{0.5cm}0.15\\
&\hspace{0.5cm}Sm&\hspace{1cm}103&\hspace{0.5cm}0.15\\
\hline
\end{tabular}
$^*$ If the morphological type of a galaxy was not specified, we examined two possible scenarios (see \S~\ref{sec:GW_Strain}).
\end{table}

Next, we estimated the bulge mass as $M_{\mathrm bulge}=f_{\mathrm bulge} M_*$, with $f_{\mathrm bulge}$ the fraction of the stars that reside in the bulge. 
For early-type galaxies, we assigned their total stellar masses to their bulges ($f_{\mathrm bulge}=1$), whereas for late-type galaxies, we used the bulge-to-total mass ratio, determined from image decomposition \citep{Weinzirl2009}, which depends on the galaxy's sub-class. The bulge fraction also depends on the existence of a galactic bar \citep[e.g.,][Figure~14]{Weinzirl2009}, but, since the 2MRS catalog typically does not specify whether a galaxy has a bar, we used the average value of barred and unbarred galaxies for each sub-class. In Table~\ref{Table:Galaxy_type}, we summarize the adopted values of bulge-to-total mass ratio (or equivalently the bulge fraction $f_{\mathrm bulge}$), as a function of the galaxy type.\footnote{We note that none of the late-type galaxies was included in NANOGrav's sensitivity volume and thus the details of the assigned bulge mass do not have a significant impact on our final results.} Additionally, the galaxy types are unknown for $\sim 40$\% of the galaxies in the catalog. For those, we estimated their bulge fractions assuming two different scenarios for the galaxy type (see \S~\ref{sec:GW_Strain}).

We estimated the SMBH mass using the correlation from \citet{McConnell2013},
\begin{equation}
    \log_{10}(M_{\mathrm{BH}})=a_2 + b_2 \log_{10}(M_{\mathrm{bulge}}/10^{11} M_{\odot}) \,,
\end{equation}
with $(a_2,b_2)=(8.46,1.05)$. 
The $M_{\mathrm{BH}}$-$\sigma$ and the $M_{\mathrm{BH}}$-$M_{\mathrm{bulge}}$ correlations have intrinsic scatters of $\epsilon=0.38$ and $\epsilon=0.44$, respectively. We used these scatters as a proxy for the uncertainties in the mass estimates, but we did not include additional sources of uncertainty (e.g., due to galaxy classification, bulge fraction, etc).

\subsection{Galaxies in the NANOGrav Volume}\label{sec:GW_Strain}

\begin{figure}
    	\includegraphics[trim=1.4cm 0.5cm 0.8cm 1cm, clip,width=0.5\textwidth]{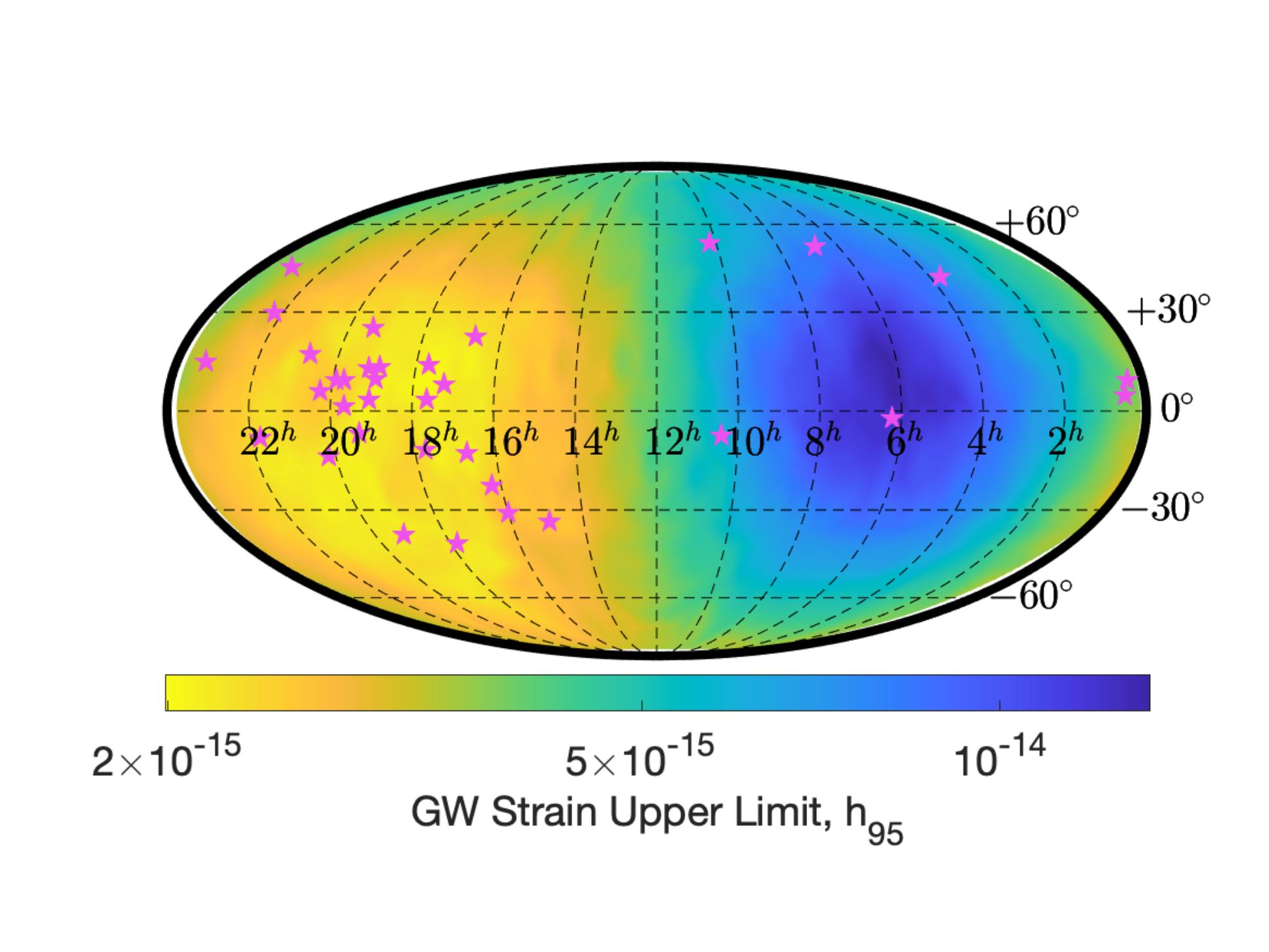}
\includegraphics[trim=1.5cm 0.5cm 0.8cm 1cm, clip,width=0.5\textwidth]{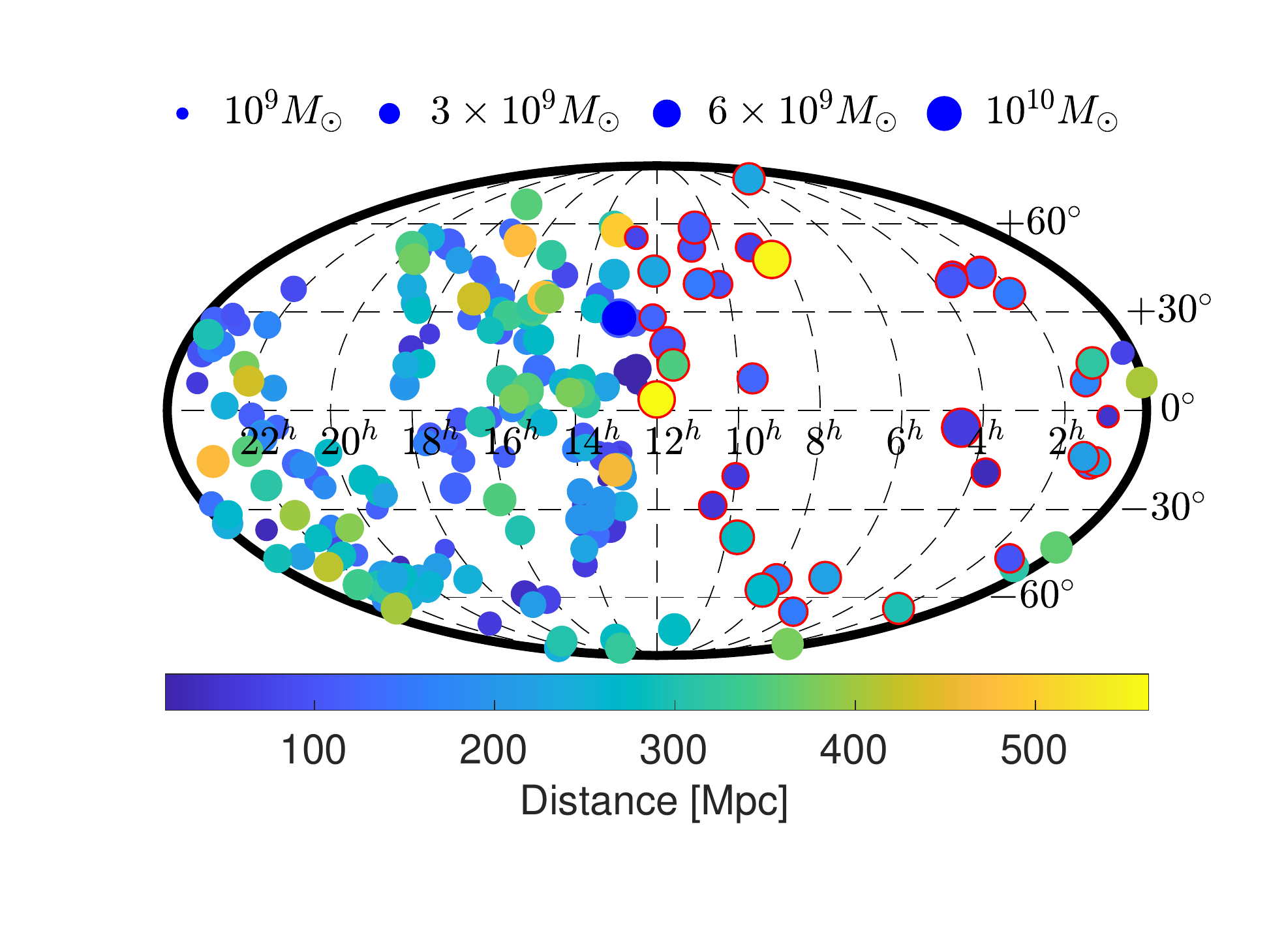}
 \caption{\emph{Top}: The 95\% upper limit on the GW strain amplitude for circular SMBHBs at $f_{\mathrm{GW}} = 8\,\mathrm{nHz}$ as a function of sky position in equatorial coordinates from the analysis of the NANOGrav 11\,yr dataset \citep{NANOGrav_CWs_2018}. The stars indicate the pulsars used in that analysis. \emph{Bottom}: Spatial distribution of 216 galaxies (194 among the galaxies of known type and 22 BCGs identified among the initially unclassified galaxies), within the NANOGrav sensitivity volume.  Colors and marker sizes reflect galaxy distances and estimated masses of the central SMBHs, respectively. Galaxies in the least sensitive half of the sky ($h_{GW}$ below the $50^{\mathrm{th}}$ percentile from the map of the top panel) are outlined with red circles.}
 \label{Fig:Early_type_galaxies}
\end{figure}

We calculated the GW strain amplitude from eq.~(\ref{eq:strain}),
using the estimates of the SMBH masses (regarded as the total masses of the binaries) and the distances obtained above. 
We considered a galaxy to be within the volume that NANOGrav can probe, if the GW strain of the tentative binary in the most optimistic scenario in terms of mass ratio and binary orbit, i.e.,
equal-mass binary in a circular orbit emitting GWs at NANOGrav's most sensitive frequency,  $h_{\mathrm{GW}}\left(\mathrm{q=1,\, f_{GW}=8\,nHz}\right)$, was equal to or higher than the upper limit from the 11\,yr dataset. However, as shown in Fig.~\ref{Fig:Early_type_galaxies} (top panel), the sensitivity of NANOGrav depends on the spatial distribution of pulsars on the sky and is highly anisotropic. We thus compared the GW strain to the upper limit in the direction of the source, $h_{95}\left(\mathrm{Ra,\,Dec}\right)$, using a skymap with resolution of 768 pixels (\textsc{healpix} with \textsc{nside=8}) from \citet{NANOGrav_CWs_2018}.  Therefore, a binary could be detectable by NANOGrav, if
\begin{equation}
    h_{\mathrm{GW}}\left(\mathrm{q=1,\, f_{GW}=8\,nHz}\right)\geq h_{95}\left(\mathrm{Ra,\,Dec}\right)
\end{equation}
In Table~\ref{Table:Galaxy_catalog}, we present the entire 2MRS galaxy catalog, with the estimated SMBH masses and galaxy distances (including the method with which each quantity was calculated) and the relevant GW upper limits at $f_{\mathrm{GW}}=$ 8\,nHz.

Among the galaxies with known types, there are 194 galaxies with SMBH mass of $10^9 - 10^{10}\,M_{\odot}$ in the volume NANOGrav can probe (up to 500\,Mpc).\footnote{NANOGrav can detect binaries with total mass of $10^9-10^{10} M_{\odot}$ up to an average distance of $\sim 15-725$\,Mpc, respectively.} The vast majority of those are early-type galaxies with only four classified as S0/a.
Next, we examined the unclassified galaxies, and estimated their masses examining two possible scenarios for the galaxy type: (1) We assumed that they are late-type Sa galaxies with $f_{\mathrm bulge}=0.31$, which corresponds to the most massive SMBHs for the case of spiral galaxies. Under this assumption, none of them are within NANOGrav's reach. (2) We assumed that they are early-type galaxies ($f_{\mathrm bulge}=1$), in which case, another 84 galaxies enter NANOGrav's sensitivity volume. We visually inspected those 84 galaxies, using data from the online database Simbad.\footnote{\url{http://simbad.u-strasbg.fr/simbad/}} We found that 22 of those are indeed early-type galaxies, classified as brightest cluster galaxies (BCGs). Six are classified as \hbox{quasars/AGN}, which we excluded because we cannot reliably estimate the stellar mass from the $K$-band magnitude (\S~\ref{sec:BH_mass}). The remaining 56 galaxies are impossible to classify based on the available images.
We kept these galaxies as a separate sub-sample, the analysis of which is presented in the Appendix.

Figure~\ref{Fig:Early_type_galaxies} (bottom panel) shows the 194 galaxies, along with the 22 BCGs among the subset with initially unknown type, that are within NANOGrav's sensitivity volume, color-coded according to their distance with the marker size denoting the estimated SMBH mass. As expected, there are significantly more galaxies in the most sensitive half of the sky, where most of the systematically monitored pulsars are located.  In the least sensitive half of the sky, binaries can be detectable only if they reside in very massive or in relatively nearby galaxies. 

Additionally, we calculated the number of galaxies that NANOGrav can probe considering the uncertainty in the mass estimates. Assuming the total mass of the binary to be equal to $M_{\mathrm{BH}}-\sigma$ and $M_{\mathrm{BH}} + \sigma$, we found 20 and 2,313 galaxies with known morphological classification within the sensitivity volume. When we also included the galaxies of unknown type, there are still 20 galaxies in the volume for the lower mass bound, whereas for the high mass bound the number becomes 2,338 and 4,545 for the two scenarios described above (i.e. late vs early type galaxies). However, in this case, it is impractical to visually examine over 2,000 galaxies to verify whether they are indeed elliptical galaxies. The impact of the mass uncertainty is obvious from the large discrepancy between the above numbers. We further explore this issue in \S~\ref{sec:mass_uncertainty}.

\begin{table}
\caption{2MRS galaxy catalog with estimates of the SMBH mass and distance, along with the GW strain upper limit towards the direction of each galaxy.}\label{Table:Galaxy_catalog}
\centering
\begin{tabular}{lcccccc}
\hline
2MASS Name$^*$               & Dist & Mass & GW   & \multicolumn{2}{c}{Method$^{**}$}\\
 &      &      & Limit& &\\
                      &[Mpc]&[$M_{\odot}$]&[$\times 10^{-15}$]& &\\
\hline
\hline
\noalign{}
J00424433$+$4116074 & 0.82 & 8.15 & 4.22 & 1 & 1\\
J00473313$-$2517196 & 3.81 & 7.74 & 3.76 & 1 & 4\\
J09553318$+$6903549 & 3.87 & 7.90 & 5.51 & 1 & 1\\
J13252775$-$4301073 & 3.92 & 7.77 & 3.31 & 1 & 1\\
J13052727$-$4928044 & 3.99 & 7.22 & 3.28 & 1 &  4\\
...&...&...&...&...\\
\end{tabular}
\flushleft
$^*$JHHMMSSss$\pm$DDMMSSs\\
$^{**}$Method used to calculate the distance and the SMBH mass. See online catalog for a detailed description.

\end{table}

\section{GW search methodology}
\label{sec:GW_Analysis}



As shown in Fig.~\ref{Fig:Early_type_galaxies}, the GW signal induced by a SMBHB on a \hbox{PTA} depends on the position of the source with respect to the position of the pulsars in the array.  
For this reason, we derived the GW upper limits towards the direction of each galaxy. In particular, we modeled the timing residuals $\delta t$ as
 \begin{equation}
     \delta t = M \epsilon + n_\mathrm{white} + n_\mathrm{red} + s \,,
     \label{eq:dt}
 \end{equation}
where $M \epsilon$ represents the timing model (with $M$ the design matrix for the linearized timing model, and
$\epsilon$ a vector of timing model parameters),
$n_\mathrm{white}$ and
$n_\mathrm{red}$ describes the white and red noise of the pulsars, respectively, 
and $s$ the GW signal, which depends on the dimensionless GW strain amplitude from eq.~(\ref{eq:strain}), the GW polarization, the source location (i.e., the so-called antenna-pattern), etc. Details on the analysis can be found in \citet{NANOGrav_CWs_2018} and \citet{Multimessenger_NANOGrav}. 

We computed the upper limits performing a Bayesian analysis with fixed GW frequency, and repeated the analysis for 11 distinct frequencies linearly spaced in the interval $[2.75,10]$\,nHz. Each Bayesian analysis requires $\sim10^3$ CPU Hours due to the large number of parameters that must be included in the model of eq.~(\ref{eq:dt}). Therefore, in order to reduce the computational cost, instead of analyzing each galaxy individually, we divided the sky into 192 pixels  of equal area (\textsc{healpix} with \textsc{nside=4}) and conducted the analysis only for the 110 pixels containing galaxies in our sample. Nearby pixels have very similar upper limits, and thus performing an upper limit analysis for each galaxy individually would be significantly more expensive computationally and would provide only incremental improvements.


\begin{figure}
    \includegraphics[width=0.5\textwidth]{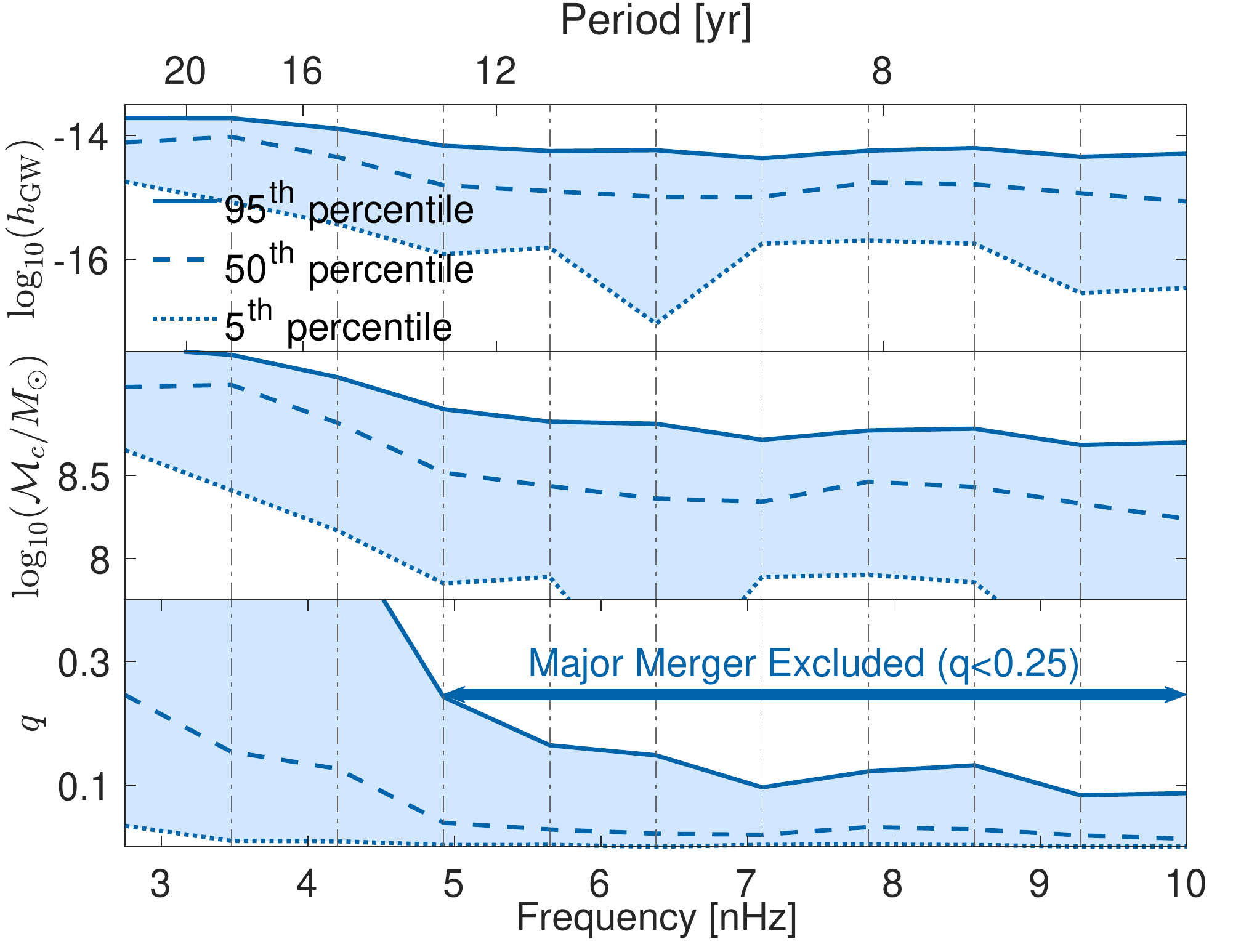}
    \caption{Illustration of the 95\% upper limits on the GW strain (top panel), the chirp mass (middle panel) and the mass ratio (bottom panel) as a function of GW frequency and orbital period for a circular SMBHB for the galaxy M49.}
    \label{fig:upper_limit_illustration}
\end{figure}

Using the upper limits of the relevant pixel, we then calculated the constraints on hypothetical binaries as a function of the GW frequency, or equivalently the orbital period of a circular binary. In Fig.~\ref{fig:upper_limit_illustration}, we illustrate the steps of the analysis for the galaxy M49 (or 2MASS J12294679$+$0800014) in our sample, an early type galaxy at a distance of 17.2\,Mpc with a SMBH mass of $2.4\times10^9 M_{\odot}$. The top panel shows the distribution ($5^{\mathrm{th}}$ to $95^{\mathrm{th}}$ percentile) of the GW strain, $h_{\mathrm{GW}}$. Using eq.~(\ref{eq:strain}), and the galaxy distance we calculated in \S~\ref{sec:dist}, we converted the distribution (and in turn the upper limits) of $h_{\mathrm{GW}}$ to the distribution of the chirp mass, which we present in the middle panel of Fig.~\ref{fig:upper_limit_illustration}. We then solved eq.~(\ref{eq:chirp_mass}) for the mass ratio using the mean SMBH mass from \S~\ref{sec:BH_mass} as the total mass of the binary
\begin{equation}
\label{eq:q}
    q=\frac{1-\sqrt{1-4 (\mc/M_{\mathrm{tot}})^{5/3}}}{1+\sqrt{1-4 (\mc/M_{\mathrm{tot}})^{5/3}}}
\end{equation}
This provided the distribution of the mass ratio as a function of frequency, which is shown in the bottom panel of the figure. Next, we examined the existence binaries delivered by major galaxy mergers with mass ratio at least 1/4. For the range of NANOGrav frequencies where the derived mass ratio upper limit was less than $1/4$, such binaries can be excluded (see \S~\ref{sec:major}). This range is indicated with an arrow at the bottom panel of Fig.~\ref{fig:upper_limit_illustration}. 
The distinct frequencies we analyzed are indicated by vertical gray lines in the figure.

In most cases, we modeled the timing residuals from eq.~(\ref{eq:dt}) sampling the GW signal directly in GW strain $h_{\mathrm{GW}}$. The uncertainty in the mass estimates was not included in this analysis.
However, for a small subset of galaxies (which host the binaries with the highest S/N), we expressed the GW signal in eq.~(\ref{eq:dt}) in terms of mass ratio and sampled directly in $q$, using the distribution of the total mass (with its uncertainty) as a prior. The reason for limiting this analysis to only five galaxies is again computational efficiency.


\section{Results}\label{sec:results}
\subsection{Galaxy Ranking based on signal-to-noise ratio}
Identical binaries (in terms of total mass, mass ratio, and distance) may have different probabilities of detection depending on their coordinates (Fig.~\ref{Fig:Early_type_galaxies}).  We quantified the probabilities of detection by calculating the S/N for putative binaries in the 216 galaxies that fall within the sensitivity volume of NANOGrav.  (The Appendix presents the S/N of the 56 galaxies of unknown type that are detectable by NANOGrav, only under the assumption that they are early type galaxies.)

We calculated the total S/N of a binary, adding the S/N$_j$ in each individual pulsar,
following \citet{Rosado2015}.
%
For this purpose, we assumed equal mass binaries in circular orbits, and a fixed GW frequency of 8\,nHz. Since the S/N depends on the inclination $i$, the initial phase $\Phi_0$, and the GW polarization angle $\psi$, we randomly draw these parameters from uniform distributions, in the range $[-1,1]$ for $\cos i$, $[0,\pi]$ for $\psi$ and $[0, 2\pi]$ for $\Phi_0$. We estimated the average total S/N from 1000 realizations for each putative binary. In Table~\ref{Table:All_Mass_ratios}, we ranked the binaries from the highest to lowest based on their average S/N. 

\subsection{Constraints on individual galaxies}\label{sec:individual_galaxies}
We used the NANOGrav 11\,yr dataset to derive constraints on hypothetical SMBHBs in the 216 galaxies identified in \S~\ref{sec:catalog}. 
Even though this dataset allows us to probe frequencies up to $\sim$317.8\,nHz, we limited our analysis to 10\,nHz for computational efficiency. We further explain this choice towards the end of the section. 
As mentioned above, in order to avoid unnecessary computations, we calculated the GW strain upper limits in 110 pixels (out of the 192 in which we divided the sky). 

\begin{figure}
    \includegraphics[width=0.5\textwidth]{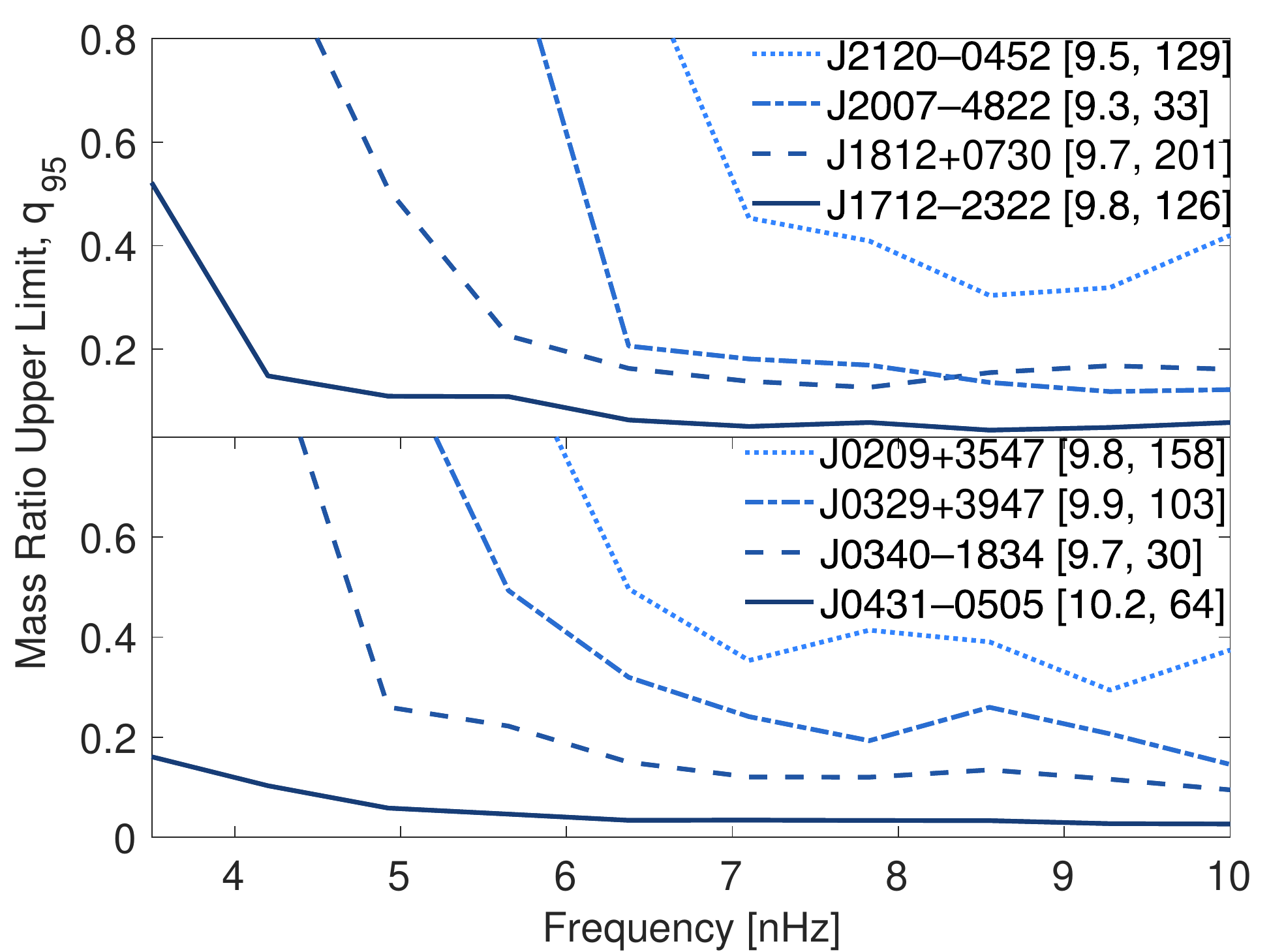}
    \caption{Constraints on the mass ratio versus GW frequency (or orbital period) for galaxies, randomly selected from the most sensitive part of the sky (top panel) and the least sensitive part of the sky (bottom panel). The abbreviated galaxy names, along with their properties (logarithmic SMBH mass in solar masses and distance in Mpc) are given in the legend.}
    \label{fig:q_examples}
\end{figure}

In Fig.~\ref{fig:q_examples}, we show the 95\% upper limits on the mass ratio (which corresponds to the solid line in the bottom panel of Fig.~\ref{fig:upper_limit_illustration}) as a function of GW frequency for eight 
galaxies in the sample. The four galaxies on the top and bottom panels are distributed close to the most and least sensitive sky location, respectively. We randomly selected these galaxies, in order to illustrate a wide range of total masses and distances of the hypothetical binaries; these quantities are shown in the brackets next to the abbreviated galaxy names in the legend.
For a significant fraction of the galaxies we can place stringent constraints on the mass ratio, for several GW frequencies. Therefore, if there is a circular binary in the center of these galaxies, the mass of the secondary SMBH must be only a few percent that of the primary. We tabulate the constraints on mass ratio in Table~\ref{Table:All_Mass_ratios}.

\begin{table*}
\caption{The 95\% mass ratio upper limits as a function of frequency for the galaxies in the NANOGrav volume ranked by the mean S/N of putative equal-mass circular binaries at 8\,nHz. The full table is available online.}\label{Table:All_Mass_ratios}
\hspace{-2.8cm}
\begin{tabular}{llllccccccccccc}
\hline
2MASS Name$^*$&Mass&Dist&S/N&$q_{95}$&$q_{95}$&$q_{95}$&$q_{95}$&$q_{95}$&$q_{95}$&$q_{95}$&$q_{95}$&$q_{95}$&$q_{95}$&$q_{95}$\\
&[$M_{\odot}$]&[Mpc]&&[2.75]$^{**}$&[3.47]&[4.20]&[4.92]&[5.65]&[6.37]&[7.10]&[7.82]&[8.55]&[9.27]&[10.0]\\
\hline
\hline
J13000809$+$2758372&10.32&112.2&240.1&0.15&0.06&0.05&0.04&0.02&0.01&0.02&0.01&0.01&0.01&0.01\\
J12304942$+$1223279&9.79&17.7&188.8&0.20&0.16&0.08&0.04&0.03&0.03&0.02&0.02&0.02&0.02&0.02\\
J04313985$-$0505099&10.23&63.8&135.9&0.42&0.16&0.10&0.06&0.05&0.03&0.03&0.03&0.03&0.03&0.03\\
J12434000$+$1133093&9.67&18.6&115.4&1.00&0.21&0.13&0.07&0.05&0.04&0.03&0.02&0.02&0.02&0.02\\
J13182362$-$3527311&9.89&53.4&99.9&0.55&0.25&0.09&0.04&0.04&0.05&0.04&0.02&0.03&0.02&0.02\\
...&...&...&...&...&...&...&...&...&...&...&...&...&...&...\\

\end{tabular}
\flushleft
$^*$JHHMMSSss$\pm$DDMMSSs. $^{**}$ GW frequencies in nHz.
\end{table*}

It is obvious from Figs.~\ref{fig:upper_limit_illustration}, and \ref{fig:q_examples} that the mass ratio upper limit is a declining function of the frequency at low frequencies, whereas at high frequencies (beyond a critical frequency), the function becomes flat and the constraints on the mass ratio are independent of frequency. The flattening of the mass ratio upper limit at high frequencies can be derived analytically, since for frequencies above 8\,nHz, the GW strain upper limit has a power-law dependence on frequency $h_{95}(f)\propto f^{2/3}$ (see \citealt{NANOGrav_CWs_2018}). From eq.~(\ref{eq:strain}) it then follows that the chirp mass upper limit for f$>$8\,nHz is independent of the frequency, $\mathcal{M}_{95}(f)=\mathrm{Const}$, and thus from eq.~(\ref{eq:q}) the mass ratio upper limit is also constant for high frequencies, i.e.,  $q_{95}(f)=\mathrm{Const}$.
This is numerically demonstrated in Fig.~\ref{Fig:q_sampling}, where we extended the analysis to higher frequencies.
As a result, our conclusions can be extended to higher frequencies without additional computations, which explains our decision to restrict the analysis to 10\,nHz.

\begin{figure}
 \centering
 \includegraphics[width=0.5\textwidth]{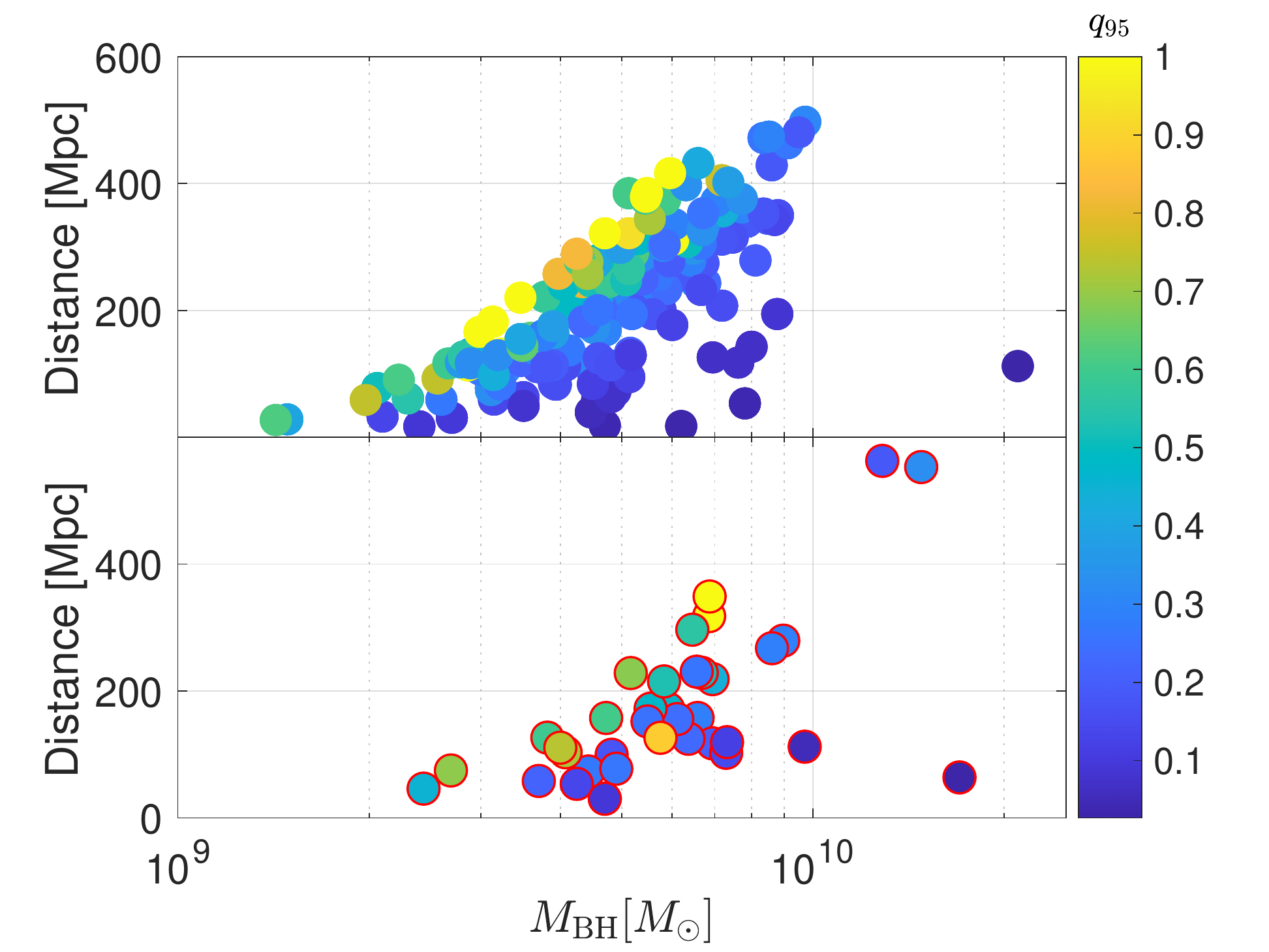}	\caption{Galaxy total mass vs galaxy distance color-coded by 95\% upper limit on the mass ratio at $f_{\mathrm{GW}} = 8$\,nHz. The top panel shows the most sensitive half of the sky, whereas the bottom panel illustrates galaxies in the least sensitive half, which are outlined with red circles as in Fig.~\ref{Fig:Early_type_galaxies}.}
 \label{fig:q_min}
\end{figure}

Since it is not practical to show the mass ratio constraints as a function of frequency for the entire sample, in Fig.~\ref{fig:q_min} we present the 95\% upper limits for a fixed frequency of 8\,nHz for the 216 galaxies in our sample, with the top/bottom panel illustrating galaxies in the most/least sensitive half of the sky.
There is an obvious trend from yellow to dark blue points, which indicates that increasingly unequal mass binaries are required, when the distances decrease for fixed mass (vertically down) and when the total masses increases for fixed distance (horizontally to the right), as expected.
The same general trend is repeated in the least sensitive half of the sky, but the overall curve is shifted down, which means that we are limited to more massive and more nearby galaxies.
We emphasize that we selected this particular frequency ($f_{\mathrm{GW}}$=8\,nHz) for two reasons: (1) this is the most sensitive frequency of NANOGrav --- for this reason, the galaxy sample was initially selected assuming tentative binaries emitting GWs at 8\,nHz, as shown in Fig. \ref{Fig:Early_type_galaxies}, and (2) this frequency corresponds to the flat part of the mass ratio upper limit curve and thus our conclusions can be extended to higher frequencies.


\subsection{Sampling in mass ratio}\label{sec:qsample}
\begin{figure}
 \centering
 \includegraphics[width=0.5\textwidth]{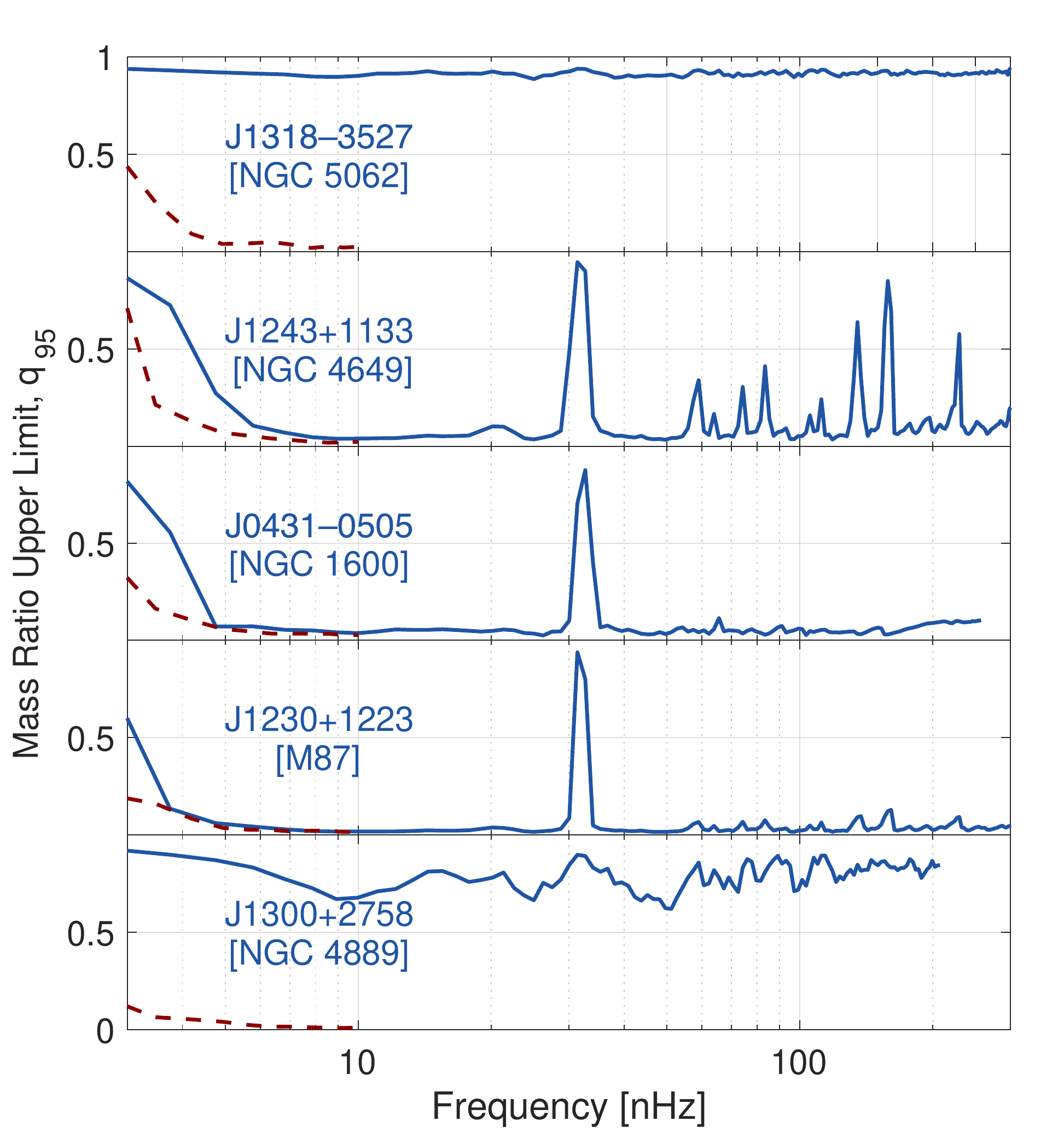}	\caption{The 95\% upper limits on mass ratio for the top five galaxies based on S/N. The blue curves show the upper limits derived by sampling the likelihood directly in mass ratio, whereas the red dashed curves show the upper limits derived from the GW strain/chirp mass upper limits, which ignore the uncertainty in the total mass.}
 \label{Fig:q_sampling}
\end{figure}

For the five top ranked galaxies based on their S/N, we run the upper limits analysis sampling the likelihood of eq.~(\ref{eq:dt}) directly in the mass ratio. This allows us to incorporate the uncertainty in the total binary mass as a prior. In Fig.~\ref{Fig:q_sampling}, we show the 95\% mass ratio upper limits as a function of frequency. For these five galaxies, we run the analysis for the full range of NANOGrav frequencies up to the cutoff frequency, which depends on the binary total mass.\footnote{For the most massive binaries, the highest frequencies NANOGrav can probe are not relevant; if we set the cut-off frequency at the respective innermost stable circular orbit (ISCO), $R_{\mathrm{ISCO}}=6 G M_{\mathrm{tot}}/c^2$, for a binary with total mass $\log_{10}(M_{\mathrm{tot}}/M_{\odot})=10.32$ (like our top ranked galaxy NGC 4889), the cut-off frequency is at $\sim$220\,nHz.} This figure also confirms that the constraints are roughly constant at high frequencies, except for specific frequencies, for which the sensitivity of NANOGrav is limited (e.g., at $\sim$30\,nHz, which corresponds to a 
period of 1 year and is caused by fitting the pulsars' positions).

In Fig.~\ref{Fig:q_sampling}, we also show with red dashed lines the limits derived from the GW strain using the mean SMBH as the total mass of the binary in \S~\ref{sec:individual_galaxies}  for comparison. As expected, including the uncertainty leads to weaker upper limits, which is more pronounced for galaxies with highly uncertain mass measurements, like NGC 4889 and NGC 5062.\footnote{We note that the mass uncertainty of NGC 4889 is large, even though it was directly measured with dynamical methods. On the other hand, the large uncertainty for NGC 5062 is unsurprising, since its mass was estimated from the $M_{\mathrm{BH}}$-$\sigma$ relationship, which has significant intrinsic scatter.}
The other three galaxies have dynamically measured SMBH masses and thus relatively small uncertainties. The upper limits here are slightly higher, but comparable to the ones obtained above, especially at higher frequencies (flat part of the upper limit curve). Additionally, the constraints we derive for putative binaries in these three galaxies are significant with only very unequal-mass binaries allowed, i.e., the secondary SMBH should be at least 10 times less massive than the primary for the majority of frequencies we examined. 
We further explore the impact of the uncertainty in the total mass in \S~\ref{sec:mass_uncertainty}.

\subsection{Constraints on major galaxy mergers}\label{sec:major}

We explored whether the galaxies in our sample could host a binary delivered by a major merger. Major mergers are considered the mergers between galaxies with similar mass (i.e., minimum mass ratio of 1/4) and play an important role in galaxy formation and evolution \citep{Volonteri2003,Kelley2017}. Since the mass of the central SMBH is tightly correlated with the galaxy mass, we assume that a major merger would result in a binary with a mass ratio $q>1/4$. This means that major galaxy mergers are also significant for PTAs, since they produce promising binaries for GW detection.

\begin{figure}
 \centering
 \includegraphics[trim=0cm  1cm 0.5cm 1cm, clip,width=0.5\textwidth]{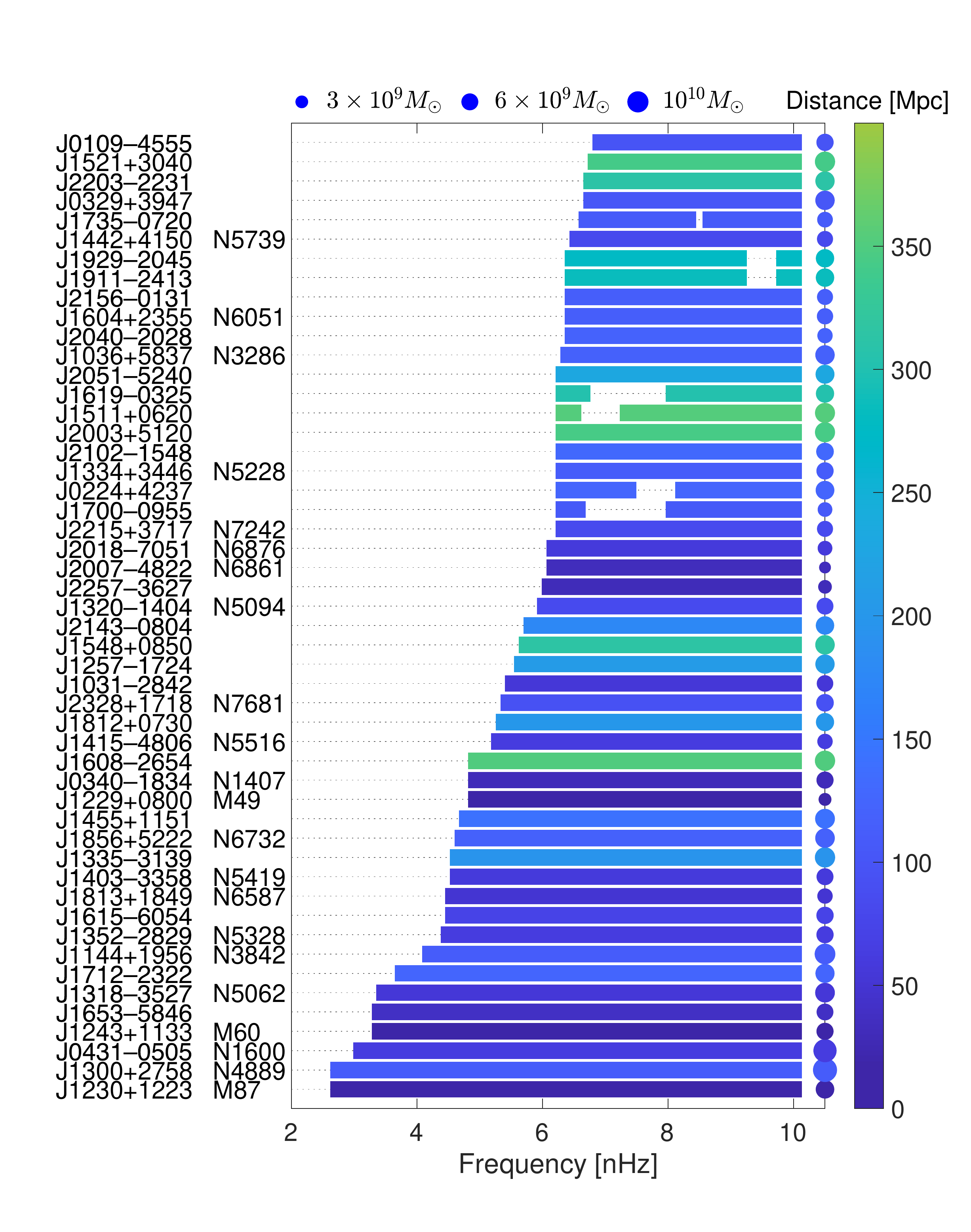}	\caption{Range of frequencies for which a binary with mass ratio $q>1/4$ delivered by a major galaxy mergers can be excluded. Each bar corresponds to a galaxy, the abbreviated coordinates of which are shown on the left. If the galaxy is included in one of the major galaxy catalogs, like the Messier or the New General Catalog, the respective names are also included. The color coding and the size of the circles on the right reflects the galaxy distance and the SMBH mass, respectively, as in Fig.~\ref{Fig:Early_type_galaxies}.}
 \label{fig:major_merger_freq}
\end{figure}

Based on the constraints on the mass ratio derived in \S~\ref{sec:individual_galaxies}, we excluded binaries delivered by major mergers for galaxies where $q_{\rm 95}<1/4$. In Fig.~\ref{fig:major_merger_freq}, we show the range of frequencies for which we were able to exclude the existence of binaries from major mergers for the 50 top-ranked galaxies based on S/N. Each bar corresponds to a distinct galaxy, the abbreviated name of which is shown on the left. The color coding reflects the distance of the binary and the size of the marker on the right the total mass of the binary (similar to Fig.~\ref{Fig:Early_type_galaxies}). It is obvious from the figure that the most informative constraints are derived for relatively nearby galaxies (within 100\,Mpc).

Motivated by the above findings, we estimated the number density of nearby galaxies, for which we can exclude binaries delivered by major mergers ($q_{\rm 95}<1/4$) at NANOGrav's most sensitive frequency ($f_{\mathrm{GW}}$=8\,nHz), as a function of the binary total mass. For this, we first estimated the maximum volume out to which NANOGrav can probe binaries from major mergers; for binaries of fixed chirp mass, i.e., fixed mass ratio (q=1/4) and total mass (see eq.~\ref{eq:strain}), the GW upper limit can be converted to a maximum distance. Due to heterogeneous sensitivity, the maximum distance is different for each pixel, and thus the volume NANOGrav can probe is a shell of irregular shape. For each pixel, we calculated the maximum distance and the respective spherical volume that NANOGrav would cover if the sensitivity was homogeneous. Since the pixels have equal area, we approximated the NANOGrav volume with a sphere of an effective volume equal to the average spherical volumes that correspond to each pixel. 
Then, we calculated the number of galaxies within this volume. 
Dividing the number of galaxies with the effective volume of the shell we get the NANOGrav limit on the density of galaxies that could host binaries delivered by major mergers at 8\,nHz. 

\begin{figure}
 \centering
 \includegraphics[trim=1cm  0.1cm 0.1cm 0.1cm, clip,width=0.5\textwidth]{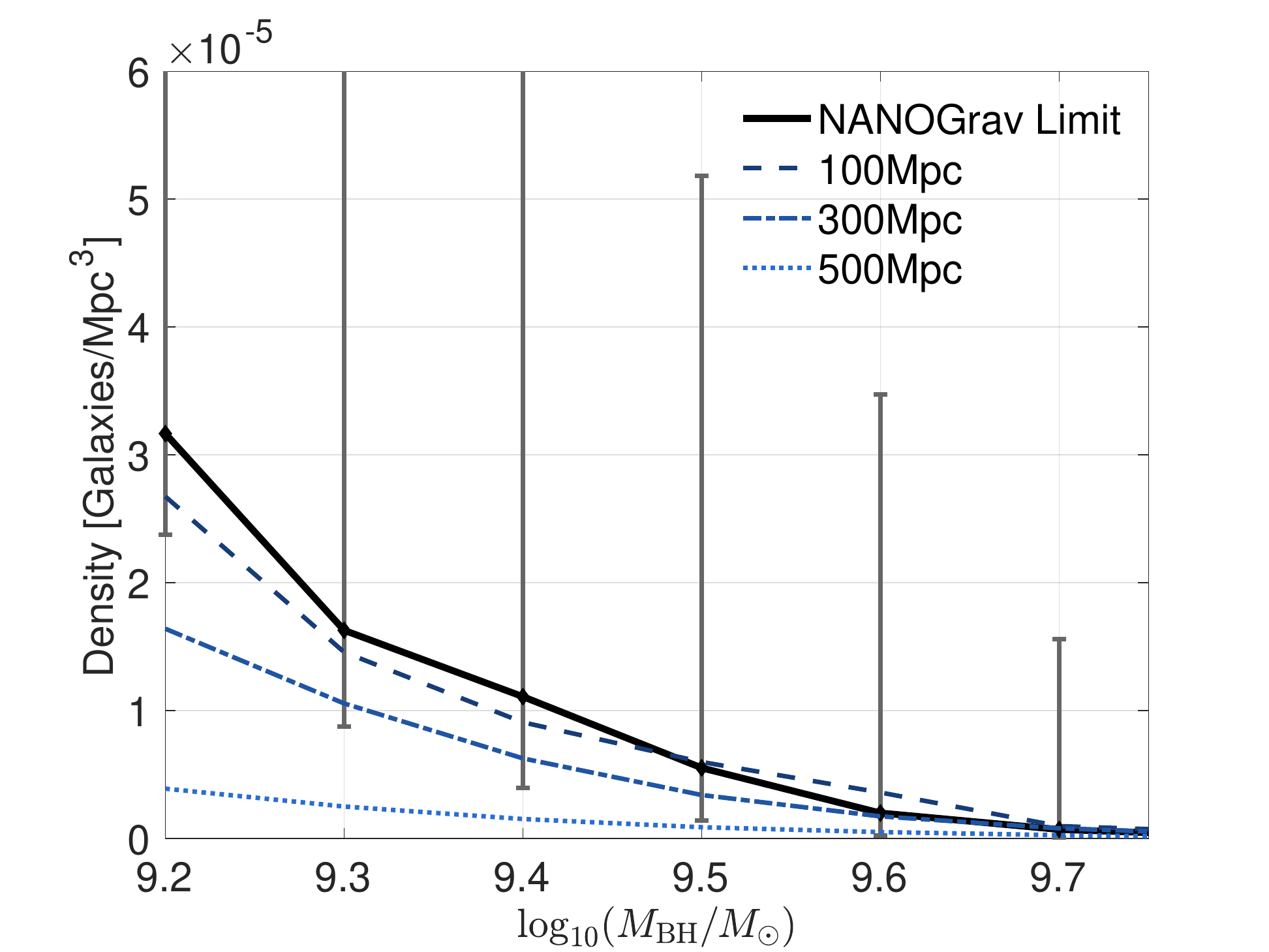}	\caption{Density of binaries delivered by major galaxy mergers ($q>1/4$) as a function of total binary mass. The shaded region shows the excluded densities based on NANOGrav constraints at a frequency of 8\,nHz. The lines show the galaxy density within spherical volumes of radii 100\,Mpc (solid line), 300\,Mpc (dashed line), and 500\,Mpc (dotted line).}
 \label{fig:major_merger}
\end{figure}

Next, we examined several discrete values for the total binary mass in the range of $10^9-10^{10} M_{\odot}$, where all the galaxies identified in \S~\ref{sec:GW_Strain} lie. 
As the binary total mass increases, the distance NANOGrav can probe becomes larger, which allows more galaxies to enter the sensitivity volume. However, a counter effect is that massive galaxies, with higher SMBH masses (e.g., with $\log_{10} (M_{\mathrm BH}/M_{\odot})>9.5$) are extremely rare. In Fig.~\ref{fig:major_merger}, we show with a black solid line the NANOGrav limit on the density of binaries produced by major mergers. The errorbars denote the range of densities, if we consider the $1\sigma$ uncertainty in the SMBH mass. For comparison, we also show the overall galaxy density above a certain fixed mass for spherical volumes with radii of 100, 300 and 500\,Mpc. We see that the density of galaxies is decreasing, when we examine a larger volume (from a radius of 100\,Mpc to a radius of 500\,Mpc). This may mean that the universe has higher density locally, potentially because we consider a small number of galaxies concentrated in local clusters (Virgo Cluster at 16\,Mpc, Coma Cluster at 120\,Mpc). An alternative explanation is that the 2MRS catalog may not be complete out to 500\,Mpc, even though this scenario is unlikely, especially for the most massive galaxies that we consider here.

We emphasize that this is the first constraint on major galaxy mergers derived directly by GW data.
However, it is worth pointing out that there are several caveats directly connecting a major merger to a binary with $q>1/4$. The main caveats are summarized below: (1) The scaling relations between the SMBH and the host galaxy have significant scatter, and thus the binary mass ratio does not necessarily reflect the mass ratio of the initial galaxies. (2) The mass of each SMBH may evolve significantly in the post-merger galaxy, meaning that the binary mass ratio at the final stages may not be directly linked to the initial mass ratio of the two SMBHs.
For instance, if the secondary SMBH accretes at a higher rate, as seen in binaries embedded in circumbinary disks, the mass ratio tends to equalize \citep{Farris2014,Siwek2020}. (3) In the mass ratio upper limits, we did not include the uncertainty of the SMBH mass estimate. Therefore, the $q<1/4$ cutoff is used merely as proxy for a binary produced by a major galaxy merger.

\section{Discussion}\label{sec:discuss}

\begin{table*}
\caption{Comparison of the galaxy catalog with related previous studies.}\label{Table:Comparison}
\centering
\begin{tabular}{llll}
\hline
\noalign{}
             &\citetalias{Schutz2016}   &\citetalias{Mingarelli2017}  &Current Paper \\ \hline \hline
Sample  &185 &5,110 &43,533\\
 &\citet{Ma2014} &2MRS Catalog (early-type, &2MRS Catalog \\
 &\citet{McConnell2013} &$D<225$\,Mpc, $M_K<-25$) &(entire catalog)\\ \hline
Distance estimates & (1) Groups Catalog &(1) Groups Catalog &(1) \emph{Cosmicflows-3}\\
&& (2) Flow-field model &(2) Groups Catalog\\
&&&(3) Flow-field model\\ \hline 
SMBH mass estimates &(1) Dynamical methods &(1) Dynamical methods &(1) Dynamical methods\\
&&(2) $M_{\mathrm{BH}}$-$M_{\mathrm{bulge}}$ &(2) Reverberation mapping \\
&&&(3) $M_{\mathrm{BH}}$-$\sigma$\\
&&&(4) $M_{\mathrm{BH}}$-$M_{\mathrm{bulge}}$\\ \hline
In NANOGrav volume  &24 &106 &216\\ \hline

\end{tabular}
\end{table*}

\subsection{Comparison with \citetalias{Schutz2016}}
\label{sec:compare_S16}
\begin{figure}
 \includegraphics[width=\columnwidth]{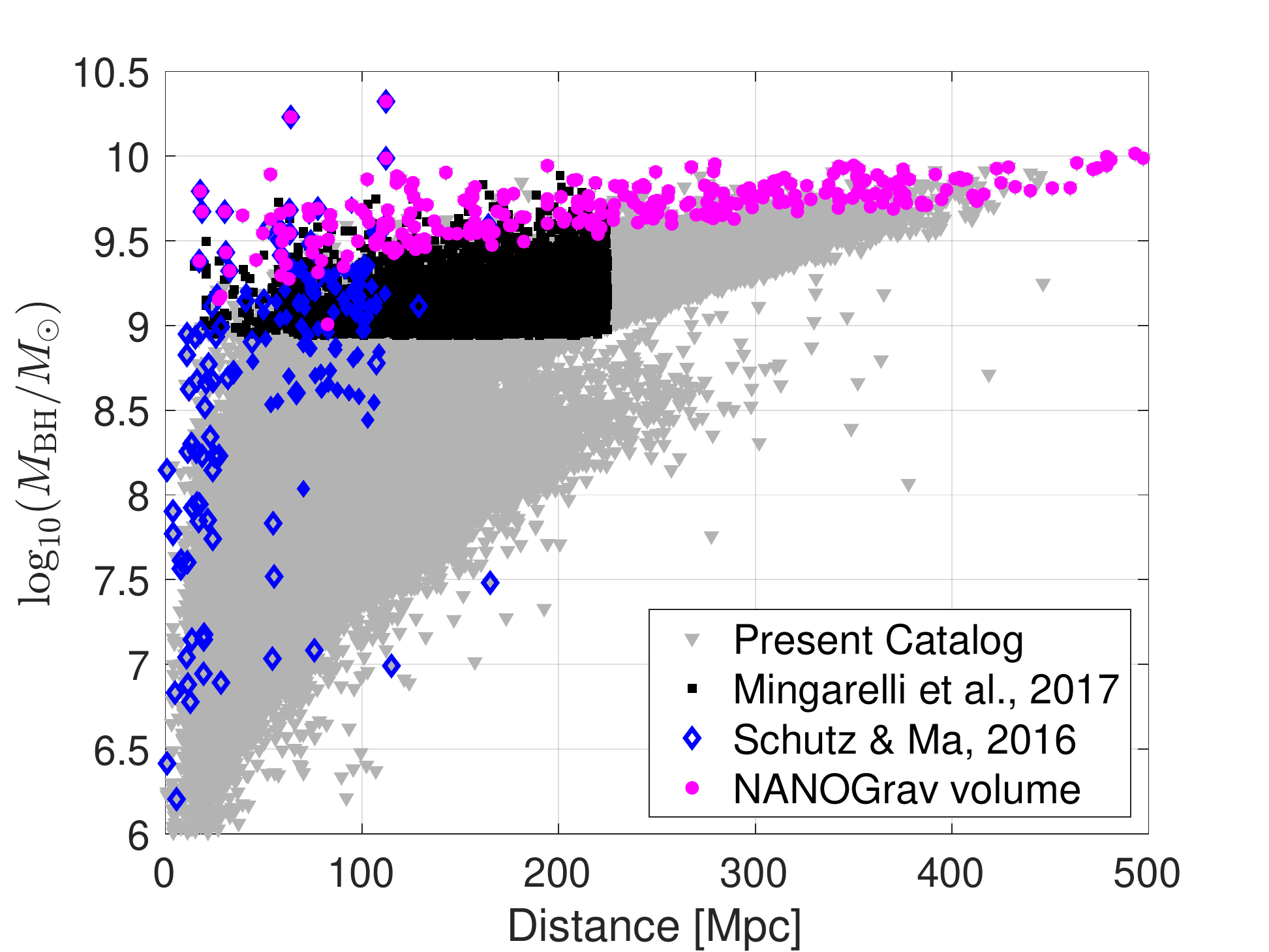}
 \caption{SMBH mass versus distance for the $\sim$45,000 galaxies in the 2MRS catalog (grey triangles), compared with the galaxy samples used in  \citetalias{Schutz2016} (blue diamonds), and in \citetalias{Mingarelli2017} (black squares). We also indicate the 216 galaxies that lie within the volume NANOGrav can probe with magenta stars.}
 \label{fig:Catalogs}
\end{figure}

\citetalias{Schutz2016} used PPTA and EPTA upper limits and performed a similar analysis on a sub-set of galaxies. In particular, they focused on galaxies with dynamical mass measurements (77 galaxies) and galaxies from the MASSIVE survey (116 galaxies). The total sample consisted of 185 galaxies, since eight of the MASSIVE galaxies had dynamical mass measurements. In Fig.~\ref{fig:Catalogs}, we show with blue diamonds the distribution of these galaxies as a function of distance and SMBH mass. We compare them with the sample of galaxies in this paper illustrated with gray triangles, and the galaxies in \citet{Mingarelli2017} shown with black squares (see below). 
We also summarize the galaxy selection and properties, along with key differences between the samples in Table~\ref{Table:Comparison}.
Of the 185 galaxies, 33 were in the sensitivity volume of PPTA and 19 in the EPTA volume, respectively. In \S~\ref{sec:GW_Strain}, we found that 24 of these galaxies are within the NANOGrav volume at 8\,nHz. Since each array has distinct and heterogeneous sensitivity, which depends on the distribution of pulsars each collaboration systematically monitors, it is not surprising that each array can probe a different number of binaries. For instance, PPTA is more sensitive in the southern hemisphere, where most of the PPTA pulsars lie (e.g., Fig.~2 in \citetalias{Schutz2016}).

\begin{figure}
 \includegraphics[width=\columnwidth]{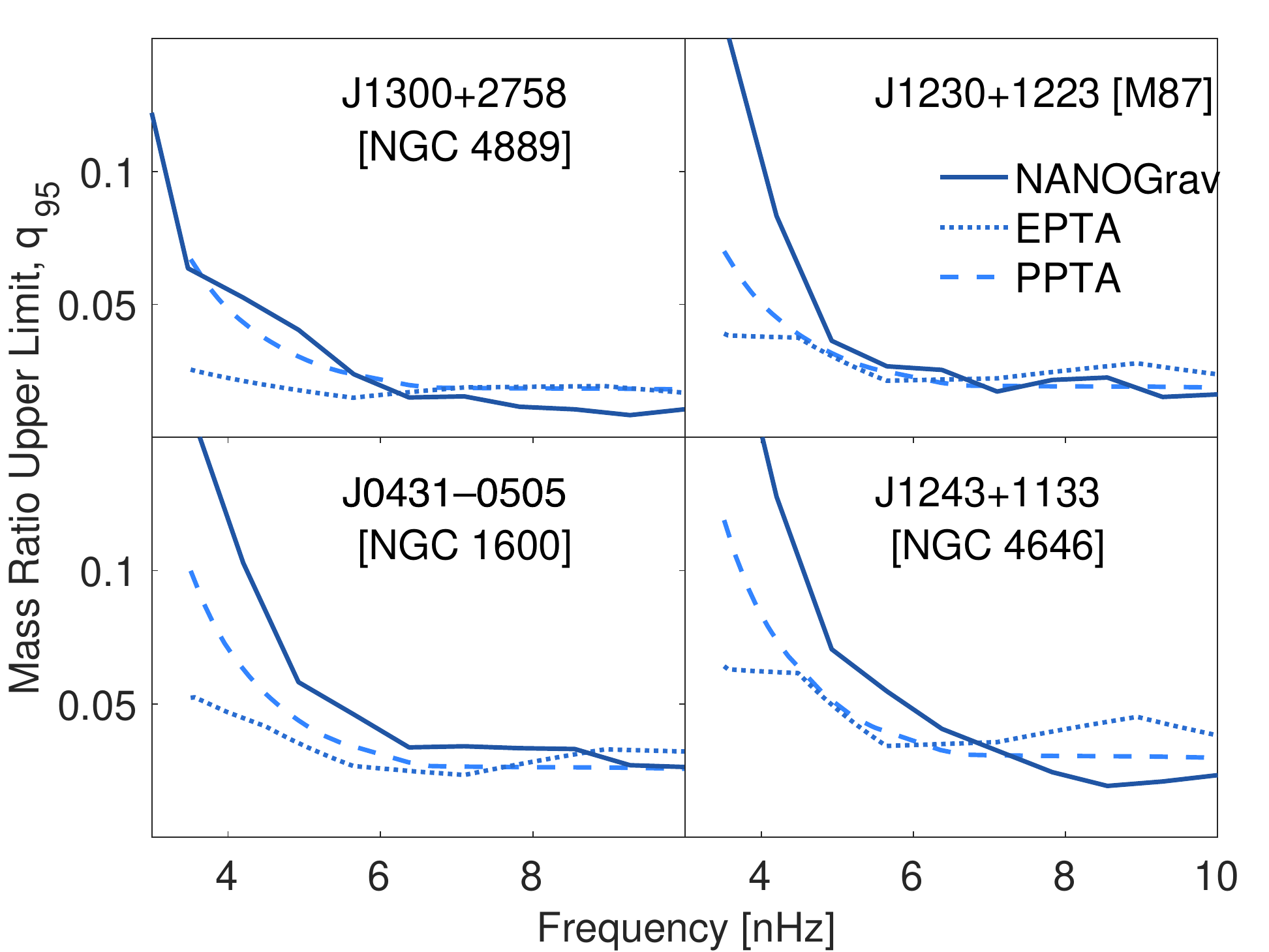}
 \caption{Mass ratio upper limits from the NANOGrav 11\,yr dataset (solid lines), compared to upper limits from PPTA (dashed lines) and EPTA (dotted lines), for four galaxies in \citetalias{Schutz2016}.}
 \label{fig:Schutz_Comparsion}
\end{figure}

In Fig.~\ref{fig:Schutz_Comparsion}, we show the mass ratio constraints for four of the top galaxies in \citetalias{Schutz2016} and compare with the 95\% upper limits we obtained in \S~\ref{sec:individual_galaxies}. We see that the NANOGrav upper limits are comparable or slightly better (e.g., NCG 4889, NGC 4649) in the flat part of the upper limit curve. On the other hand, at the lowest frequencies, our limits are less constraining. The differences between the results of the two studies are mainly due to a combination of the following reasons: 
(1) There are slight differences in the assumed properties of the putative binaries in the two studies. The total masses of the binaries in these four galaxies are identical, but the distances in our study are slightly larger. All else being equal, this would naturally lead to slightly weaker upper limits for the binaries considered here.
(2) At the most sensitive frequency of 8\,nHz, the sky averaged 95\% upper limit of NANOGrav is approximately $h_0 < 7.3 \times 10^{-15}$, which is 1.4 times lower than the EPTA limit ($h_0 < 10^{-14}$), and 2 times lower than the PPTA limit ($h_0 < 1.7 \times 10^{-14}$) used in \citetalias{Schutz2016}. The improvement in the GW limits should lead to more stringent constraints on the mass ratio and partly explains why NANOGrav performs better at higher frequencies.
(3) PPTA and EPTA have increased sensitivity at low frequencies, because some of the pulsars have long baselines of timing observations (longer than 17\,yr). Therefore, the limits derived with the NANOGrav 11\,yr dataset are less constraining at these low frequencies.
(4) The NANOGrav upper limits vary significantly from the most sensitive to the least sensitive sky location (from $h_0 < 2.0\times10^{-15}$ to $h_0 < 1.34 \times 10^{-14}$). This means that the relative position of the galaxy to the pulsars in the array is important and thus discrepancies between the derived limits from the three arrays are expected. An example of this is NGC 1600, which is located at NANOGrav's least sensitive part of the sky, where EPTA has at least four pulsars in its vicinity.

\subsection{Comparison with \citet{Mingarelli2017}}
\label{sec:compare_M17}

\citet{Mingarelli2017} [hereafter \citetalias{Mingarelli2017}] modeled the local GW sky, using galaxy merger rates from \emph{Illustris}, and the local massive galaxies as an observational anchor. For this, they extracted a sample of 5,110 massive early-type galaxies from 2MRS,\footnote{ \url{https://github.com/ChiaraMingarelli/nanohertz_GWs/blob/master/galaxy_data/2mass_galaxies.lst}} and calculated the probability of each galaxy hosting a GW-emitting SMBHB system. These galaxies are illustrated with black squares in Fig.~\ref{fig:Catalogs}. 

Even though the starting point of both studies is the 2MRS catalog, the goals of the two studies are complementary, which explains why different selection criteria and cuts were employed.
For instance, \citetalias{Mingarelli2017} assessed the likelihood of PTAs detecting GWs from individual sources, through multiple realizations of the local universe and thus the completeness of the galaxy sample is important. On the other hand, we present a purely observational approach, in which we used the current GW upper limits to place constraints on circular binaries. For our purposes, compiling the most comprehensive catalog that includes all the galaxies that NANOGrav can probe is crucial.

Examining the galaxies in the \citetalias{Mingarelli2017} catalog,
we found that only 106 lie in the NANOGrav sensitivity volume, a factor of two lower compared to the 216 from the catalog that we use here. This discrepancy can be understood in light of the selection criteria and calculation of the galaxy properties (SMBH masses, and distances). Below we summarize the key differences between the catalogs (but see also Table~\ref{Table:Comparison}):

\begin{enumerate}
    \item We examined the entire 2MRS catalog, whereas \citetalias{Mingarelli2017} imposed cuts on the absolute $K$-band magnitude ($M_K<-25$) and the galaxy distance ($D<225$\,Mpc). The magnitude cut has no effect on the sample of interest (i.e., galaxies with binaries detectable by NANOGrav), as it mostly excludes low-mass galaxies. However, the distance cut is significant, because it excludes massive galaxies at relatively large distances, but within NANOGrav's capabilities (e.g., see Fig.~\ref{fig:Catalogs}). 
    
    \item We used a multi-layered approach for the estimates of SMBH masses. The difference with \citetalias{Mingarelli2017} lies in the inclusion of mass estimates from reverberation mapping for AGN, and the estimates from the $M_{\mathrm{BH}}$-$\sigma$ relationship. Both methods provide better estimates compared to the $M_{\mathrm{BH}}$-$M_{\mathrm{bulge}}$ relationship, because the former provides dynamical mass measurements, whereas the latter relies on fewer assumptions (see also \S~\ref{sec:mass_uncertainty}).\footnote{There are several steps involved in estimating the SMBH mass from the $M_{\mathrm{BH}}$-$M_{\mathrm{bulge}}$ relationship (estimation of stellar mass from the $K$-band magnitude, estimation of the bulge fraction, etc), which may add significant uncertainty.}

    \item For the distance estimates, we started from the Cosmicflows-3 catalog, since it compiles the highest-quality distance measurements available in the literature. For galaxies not included in Cosmicflows-3, we used the catalog of galaxy groups \citep{Crook2007} or corrected for the local velocity field \citep{Mould2000}, as in \citetalias{Mingarelli2017}. For the latter subset, the distances in the two catalogs are in excellent agreement, with the exception of galaxies in the vicinity of the Great Attractor, to which we assigned the velocity of the cluster. For the subset of galaxies in Cosmicflows-3, the distances in \citetalias{Mingarelli2017} are overestimated. 

    \item We examined the entire 2MRS catalog, whereas \citetalias{Mingarelli2017} selected early-type (elliptical and S0) galaxies. Additionally, the classification in \citetalias{Mingarelli2017} was based on the extragalactic database Hyperleda, whereas we used the classification from the 2MRS catalog itself. We note that a comparison of galaxy types in the two catalogs indicates that not all 5,110 galaxies in \citetalias{Mingarelli2017} are classified as early-type in the 2MRS classification; a fraction of them ($\sim$16\%) are classified as late-type galaxies, whereas $\sim$15\% are unclassified. With visual inspection of images in Simbad, we confirmed that at least some of the galaxies in the \citetalias{Mingarelli2017} sample are indeed late-type (spiral) galaxies. Although it is impossible to determine whether this affects their overall conclusions, if spiral galaxies are misclassified as elliptical, the general trend is that the assumed SMBH mass and in turn the GW strain are overestimated.

\end{enumerate}

\subsection{Uncertainty in total mass}\label{sec:mass_uncertainty}
This work presented a study on how the NANOGrav data can be used to place constraints on SMBHBs in massive nearby galaxies. For most galaxies, we derived the chirp mass upper limits from the GW strain, and converted them to mass ratio upper limits using the mean SMBH mass (measured or estimated from scaling relations) as the total mass of the binary. As explained above, the main reason for this approach is computational efficiency. However, one of the caveats is that it ignores the uncertainty in the total binary mass. As we demonstrated in \S~\ref{sec:qsample}, incorporating the SMBH mass uncertainty deteriorates the mass ratio upper limits, with larger uncertainties having a stronger impact on these constraints. 

\begin{figure}
 \includegraphics[width=\columnwidth]{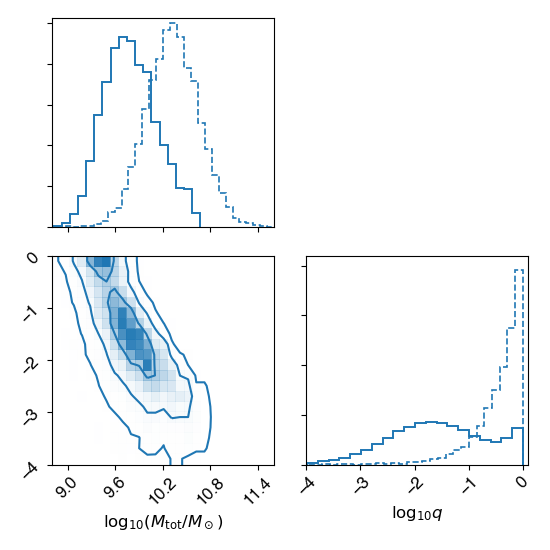}
 \caption{Posterior distribution of total mass and mass ratio for galaxy NGC 4889 obtained in \S~\ref{sec:qsample}. The prior distribution of the total mass is also shown with the dotted histogram.}
 \label{fig:corner}
\end{figure}

Here we further explore the effect of the mass uncertainty on the mass ratio upper limits. For this, we derived the upper limits from the GW strain (as in \S~\ref{sec:individual_galaxies}), but instead of using the mean SMBH mass in eq.~(\ref{eq:q}), we sampled the total mass from the entire mass distribution, which includes the uncertainty. Next, we compared with the upper limits we obtained sampling directly in the mass ratio in \S~\ref{sec:qsample}, where the mass uncertainty was included as a prior in the analysis. We find that when the uncertainty is small (e.g., NGC 4649, NGC 1600, M87), the upper limits from the different methods are in excellent agreement. However, if the mass uncertainty is large (e.g., NGC 5062, NGC 4889), the mass ratio upper limits tend to be higher when we sample directly in mass ratio, which means that the two methods are not always equivalent. We carefully examined the cause for this discrepancy for the galaxy NGC 4889 at a fixed frequency of 8\,nHz, which we illustrate in Fig.~\ref{fig:corner}. We see that the posterior distribution of the total mass is skewed towards smaller values compared to the prior, which is shown with a dotted histogram. This results in a tail in the mass ratio distribution,  which leads to significantly higher 95\% mass ratio upper limits.

Currently, this is a significant limitation, since sampling the mass ratio for each galaxy is computationally expensive. This is why we chose to perform this type of analysis only for five tentative binaries. We defer a similar analysis for the entire sample to a future study. Additionally, for the vast majority of galaxies, the SMBH mass (attributed to the total mass of the binary) was estimated from global scaling relationships, the intrinsic scatter of which inevitably results in large uncertainties. The uncertainty is small only for galaxies with precise dynamical mass measurements, which are observationally demanding and limited to a small number of galaxies. Such observations are unlikely to become readily available for a large number of galaxies in the near-future. However, in this study, we also included a small subset of AGN with SMBH masses measured from reverberation mapping. This is a promising direction, since planned multi-epoch spectroscopic surveys, like SDSS-V \citep{Kollmeier2017}, will soon provide SMBH mass estimates for a large number of AGN through reverberation mapping. This is a particularly exciting development for future studies, as AGN are promising candidates for hosting SMBHBs.

\subsection{Additional assumptions and limitations}
The mass ratio upper limits depend on the total mass of the binary (eq.~\ref{eq:q}), which was assumed to be equal to the mass of the SMBH and for the vast majority of galaxies was calculated from scaling relationships.  In \S~\ref{sec:BH_mass} and \ref{sec:GW_Strain}, we described the assumptions required to determine the SMBH mass, such as the bulge fraction, the galaxy classification, the existence of a bar, etc. We also demonstrated the challenges in determining the SMBH mass, when the galaxy type is unknown. 
Another effect is related to the choice of the SMBH mass correlations; throughout the analysis, we have adopted the scalings ($M_{\mathrm{BH}}$-$\sigma$ and $M_{\mathrm{BH}}$-$M_{\mathrm{bulge}}$) from \citet{McConnell2013}. These correlations may provide higher SMBH mass estimates (due to measurements of SMBHs at the high mass end) compared to other correlations \citep{Shankar_2016}. Adopting a shallower correlation would result in lower-mass binaries and weaker mass-ratio limits. Even though all of the above affect the determination of the mean SMBH mass (and in turn the mass ratio upper limits), they are likely negligible compared to the effect of the uncertainty due to the instrinsic scatter in the $M_{\mathrm{BH}}$-$\sigma$ and $M_{\mathrm{BH}}$-$M_{\mathrm{bulge}}$ correlations, as shown above.

Another potential source of uncertainty in estimating the mass ratio limits is related to the distance determination. In this work, we did not include the uncertainty in the distance estimates, mainly because this uncertainty is typically smaller than that of the SMBH mass (especially when obtained from scaling relations). Additionally, the dependence of the GW strain on the distance is weaker compared to the dependence on the chirp mass. Overall, even if the impact of the distance uncertainty is likely negligible, it should be incorporated in future work. Future analysis should also explore the uncertainty in the value of the Hubble constant, which is necessary for the distance calculations. 

Last but not least, in our analysis we assumed circular binaries, and thus our constraints cannot be generalized to non-circular binaries. In reality, binary orbits may deviate from circular due to high eccentricity at formation, or because the eccentricity is excited as the binary interacts with the surrounding gaseous environment \citep{2005ApJ...634..921A}. (We note, however, that GW emission tends to circularize the binary orbits.) In eccentric binaries the GW strain is split among multiple harmonics, resulting in weaker GW signal. In practice, this means that if the galaxies we analyzed hosted eccentric binaries, the limits we would derive would be less constraining that the ones we presented here for circular binaries. In a future study we will derive the upper limits examining eccentric binaries, especially since non-circular binaries may be relatively common.

\section{Summary}\label{sec:summary}
We analyzed the NANOGrav 11\,yr dataset to derive limits on tentative binaries in nearby galaxies. For this, we assembled a comprehensive sample of $\sim$44,000 galaxies, for which we estimated the distance and the mass of the central SMBH, starting from the 2MRS catalog. We found that 216 massive and relatively nearby galaxies are within the volume NANOGrav can probe, i.e., if they hosted equal-mass circular binaries emitting at 8\,nHz (the most sensitive frequency of NANOGrav), the GWs would be detectable by NANOGrav. We ranked the tentative binaries in these galaxies based on the total S/N. Subsequently, we obtained mass ratio upper limits as a function of frequency for each of those galaxies. The upper limits for several galaxies are stringent and only binaries with very unequal masses are allowed. This is comparable to constraints placed on a potential SMBHB in the center of Milky Way. However, we also demonstrated that the uncertainty in the total mass of the binary can significantly weaken the upper limits and needs to be taken into account in future studies. Additionally, we placed limits on binaries delivered by major galaxy mergers and constrained the density of such binaries at NANOGrav's most sensitive frequency, which is the first such constraint based on GW data alone.

As the PTA upper limits improve (with more pulsars being systematically monitored for longer periods of time, and employing improved statistical techniques), we will be able to place more stringent limits on GW strain. This, in turn, will improve the limits on tentative nearby binaries. In particular, the combination of data from individual PTAs in an IPTA dataset will provide not only improved upper limits, but also a more homogeneous sensitivity on the sky. The enhanced sensitivity will allow us to probe a larger volume extending beyond the local universe. As a result, a larger number of galaxies will become accessible for this type of study, while the limits for the current galaxies will be improved further. In addition, NANOGrav is planning targeted pulsar searches towards the direction of promising galaxies (e.g., promising massive elliptical galaxies, or binary candidates identified in electromagnetic searches). This is significant, since proximity to a pulsar with high quality data has the potential to improve the mass ratio upper limits. All of above will lead to an overall better understanding on the existence of SMBHBs in the local universe and beyond.

\acknowledgments
\textit{Author Contributions}: We list specific contributions to this paper below. MC led the work on this paper, compiled the galaxy catalog, derived the mass ratio upper limits and wrote the manuscript. SJV run the GW analysis. CPM, TJWL and JS contributed with significant discussions in shaping the project. NANOGrav data is the result of the work of dozens of people over the course of more than thirteen years. ZA, PBD, MED, TD, JAE, ECF, EF, PAG, MLJ, MTL, RSL, MAM, CN, DJN, TTP, SMR, PSR, RS, IHS, KS, and JKS developed the 11-year data set. All authors are key contributing members to the NANOGrav collaboration.\\
The NANOGrav project receives support from National Science Foundation (NSF) Physics Frontiers Center award number 1430284. 
MC and SRT acknowledge support from NSF, award number 2007993.
Part of this research was carried out at the Jet Propulsion Laboratory, California Institute of Technology, under a contract with the National Aeronautics and Space Administration. 
We are grateful for computational resources provided by the Leonard E.~Parker Center for Gravitation, Cosmology \& Astrophysics at the University of Wisconsin-Milwaukee, which is supported by NSF Grant PHY-1626190.
Data for this project were collected using the facilities of the Green Bank Observatory and the Arecibo Observatory. The Green Bank Observatory is a facility of the National Science Foundation operated under cooperative agreement by Associated Universities, Inc. The Arecibo Observatory is a facility of the National Science Foundation operated under cooperative agreement by the University of Central Florida in alliance with Yang Enterprises, Inc.~and Universidad Metropolitana.

\clearpage

%


\software{\texttt{enterprise} \citep{enterprise}, \texttt{PTMCMCSampler} \citep{evh17b}}



\appendix
In \S~\ref{sec:GW_Strain}, we examined whether a galaxy lies in the sensitivity volume of NANOGrav. For this, we calculated the mass of the central SMBH and assigned it to the total mass of the binary. Since we primarily used the global scaling relationship $M_{\mathrm{BH}}$-$M_{\mathrm{bulge}}$, it is necessary to identify the morphological type of the galaxy, because the fraction of the stars in the bulge depends on this. However, the galaxy classification is unknown for about 40\% of the galaxies in 2MRS. For those we examined two scenarios: (1) the galaxies are early type, so $f_{\mathrm bulge}=1$ and (2) they are late-type with $f_{\mathrm bulge}=0.31$. We found that 56 galaxies of unknown type would be detectable by NANOGrav under the assumption they are early-type galaxies. Here we present the results for this sub-set of galaxies. 

\begin{figure}
 \includegraphics[width=\columnwidth]{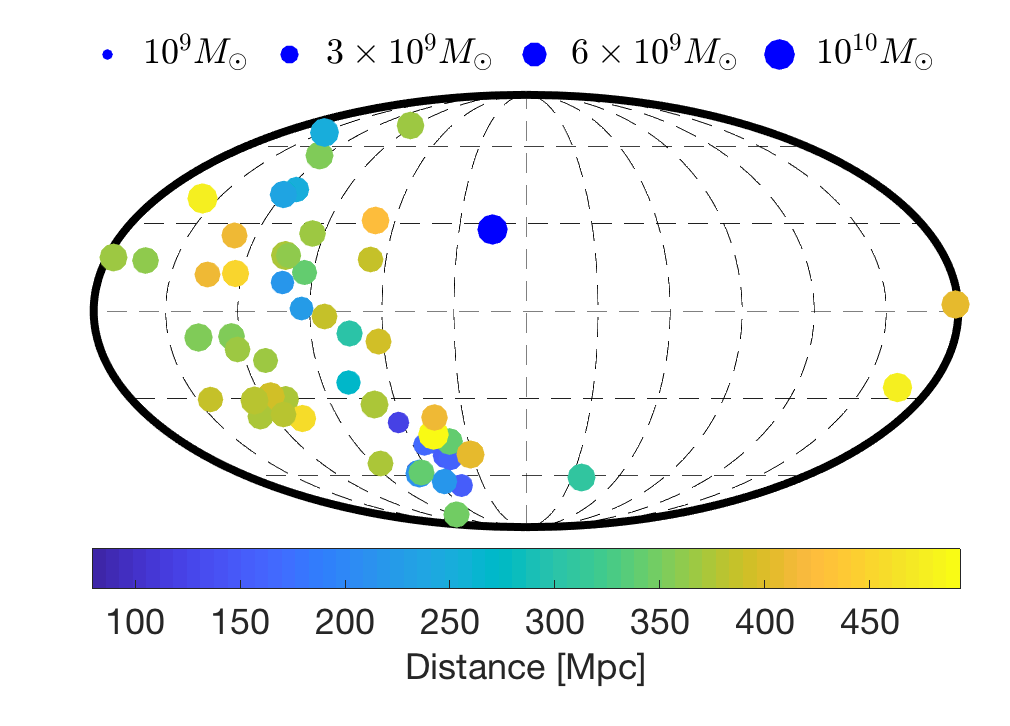}
 \caption{Spatial distribution of 56 galaxies of unknown type that could be within the NANOGrav sensitivity volume if they are elliptical galaxies.  Colors and marker sizes similar to Fig.~\ref{Fig:Early_type_galaxies}.}
 \label{fig:unkown_type}
\end{figure}

In Fig.~\ref{fig:unkown_type}, we show the distribution of these 56 galaxies on the sky. We see that these galaxies are distributed relatively close to the galactic plane, where NANOGrav's sensitivity is maximum and typically are at large distances. This could explain the scarcity of high quality images that would allow their morphological classification. 

\bibliography{galaxies}{}

\begin{thebibliography}{}
\expandafter\ifx\csname natexlab\endcsname\relax\def\natexlab#1{#1}\fi
\providecommand{\url}[1]{\href{#1}{#1}}
\providecommand{\dodoi}[1]{doi:~\href{http://doi.org/#1}{\nolinkurl{#1}}}
\providecommand{\doeprint}[1]{\href{http://ascl.net/#1}{\nolinkurl{http://ascl.net/#1}}}
\providecommand{\doarXiv}[1]{\href{https://arxiv.org/abs/#1}{\nolinkurl{https://arxiv.org/abs/#1}}}

\bibitem[{{Aggarwal} {et~al.}(2019){Aggarwal}, {Arzoumanian}, {Baker},
  {Brazier}, {Brinson}, {Brook}, {Burke-Spolaor}, {Chatterjee}, {Cordes},
  {Cornish}, {Crawford}, {Crowter}, {Cromartie}, {DeCesar}, {Demorest},
  {Dolch}, {Ellis}, {Ferdman}, {Ferrara}, {Fonseca}, {Garver-Daniels},
  {Gentile}, {Hazboun}, {Holgado}, {Huerta}, {Islo}, {Jennings}, {Jones},
  {Jones}, {Kaiser}, {Kaplan}, {Kelley}, {Key}, {Lam}, {Lazio}, {Levin},
  {Lorimer}, {Luo}, {Lynch}, {Madison}, {McLaughlin}, {McWilliams},
  {Mingarelli}, {Ng}, {Nice}, {Pennucci}, {Pol}, {Ransom}, {Ray}, {Siemens},
  {Simon}, {Spiewak}, {Stairs}, {Stinebring}, {Stovall}, {Swiggum}, {Taylor},
  {Turner}, {Vallisneri}, {van Haasteren}, {Vigeland }, {Witt}, {Zhu}, \& {(The
  NANOGrav Collaboration}}]{NANOGrav_CWs_2018}
{Aggarwal}, K., {Arzoumanian}, Z., {Baker}, P.~T., {et~al.} 2019, \apj, 880,
  116, \dodoi{10.3847/1538-4357/ab2236}

\bibitem[{{Armitage} \& {Natarajan}(2005)}]{2005ApJ...634..921A}
{Armitage}, P.~J., \& {Natarajan}, P. 2005, \apj, 634, 921,
  \dodoi{10.1086/497108}

\bibitem[{{Arzoumanian} {et~al.}(2018{\natexlab{a}}){Arzoumanian}, {Brazier},
  {Burke-Spolaor}, {Chamberlin}, {Chatterjee}, {Christy}, {Cordes}, {Cornish},
  {Crawford}, {Thankful Cromartie}, {Crowter}, {DeCesar}, {Demorest}, {Dolch},
  {Ellis}, {Ferdman}, {Ferrara}, {Fonseca}, {Garver-Daniels}, {Gentile},
  {Halmrast}, {Huerta}, {Jenet}, {Jessup}, {Jones}, {Jones}, {Kaplan}, {Lam},
  {Lazio}, {Levin}, {Lommen}, {Lorimer}, {Luo}, {Lynch}, {Madison}, {Matthews},
  {McLaughlin}, {McWilliams}, {Mingarelli}, {Ng}, {Nice}, {Pennucci}, {Ransom},
  {Ray}, {Siemens}, {Simon}, {Spiewak}, {Stairs}, {Stinebring}, {Stovall},
  {Swiggum}, {Taylor}, {Vallisneri}, {van Haasteren}, {Vigeland}, {Zhu}, \&
  {NANOGrav Collaboration}}]{NANOGrav_11yr_2018}
{Arzoumanian}, Z., {Brazier}, A., {Burke-Spolaor}, S., {et~al.}
  2018{\natexlab{a}}, The Astrophysical Journal Supplement Series, 235, 37,
  \dodoi{10.3847/1538-4365/aab5b0}

\bibitem[{{Arzoumanian} {et~al.}(2018{\natexlab{b}}){Arzoumanian}, {Baker},
  {Brazier}, {Burke-Spolaor}, {Chamberlin}, {Chatterjee}, {Christy}, {Cordes},
  {Cornish}, {Crawford}, {Thankful Cromartie}, {Crowter}, {DeCesar},
  {Demorest}, {Dolch}, {Ellis}, {Ferdman}, {Ferrara}, {Folkner}, {Fonseca},
  {Garver-Daniels}, {Gentile}, {Haas}, {Hazboun}, {Huerta}, {Islo}, {Jones},
  {Jones}, {Kaplan}, {Kaspi}, {Lam}, {Lazio}, {Levin}, {Lommen}, {Lorimer},
  {Luo}, {Lynch}, {Madison}, {McLaughlin}, {McWilliams}, {Mingarelli}, {Ng},
  {Nice}, {Park}, {Pennucci}, {Pol}, {Ransom}, {Ray}, {Rasskazov}, {Siemens},
  {Simon}, {Spiewak}, {Stairs}, {Stinebring}, {Stovall}, {Swiggum}, {Taylor},
  {Vallisneri}, {van Haasteren}, {Vigeland }, {Zhu}, \& {NANOGrav
  Collaboration}}]{NANOGrav_GWB_2018}
{Arzoumanian}, Z., {Baker}, P.~T., {Brazier}, A., {et~al.} 2018{\natexlab{b}},
  \apj, 859, 47, \dodoi{10.3847/1538-4357/aabd3b}

\bibitem[{{Arzoumanian} {et~al.}(2020){Arzoumanian}, {Baker}, {Brazier},
  {Brook}, {Burke-Spolaor}, {Becsy}, {Charisi}, {Chatterjee}, {Cordes},
  {Cornish}, {Crawford}, {Cromartie}, {DeCesar}, {Demorest}, {Dolch},
  {Elliott}, {Ellis}, {Ferrara}, {Fonseca}, {Garver-Daniels}, {Gentile},
  {Good}, {Hazboun}, {Islo}, {Jennings}, {Jones}, {Kaiser}, {Kaplan}, {Kelley},
  {Shapiro Key}, {Lam}, {Lazio}, {Luo}, {Lynch}, {Madison}, {McLaughlin},
  {Mingarelli}, {Ng}, {Nice}, {Pennucci}, {Pol}, {Ransom}, {Ray},
  {Shapiro-Albert}, {Siemens}, {Simon}, {Spiewak}, {Stairs}, {Stinebring},
  {Stovall}, {Swiggum}, {Taylor}, {Vallisneri}, {Vigeland }, \&
  {Witt}}]{Multimessenger_NANOGrav}
---. 2020, arXiv e-prints, arXiv:2005.07123.
\newblock \doarXiv{2005.07123}

\bibitem[{{Babak} {et~al.}(2016){Babak}, {Petiteau}, {Sesana}, {Brem},
  {Rosado}, {Taylor}, {Lassus}, {Hessels}, {Bassa}, {Burgay}, {Caballero},
  {Champion}, {Cognard}, {Desvignes}, {Gair}, {Guillemot}, {Janssen},
  {Karuppusamy}, {Kramer}, {Lazarus}, {Lee}, {Lentati}, {Liu}, {Mingarelli},
  {Os{\l}owski}, {Perrodin}, {Possenti}, {Purver}, {Sanidas}, {Smits},
  {Stappers}, {Theureau}, {Tiburzi}, {van Haasteren}, {Vecchio}, \&
  {Verbiest}}]{Babak2016}
{Babak}, S., {Petiteau}, A., {Sesana}, A., {et~al.} 2016, \mnras, 455, 1665,
  \dodoi{10.1093/mnras/stv2092}

\bibitem[{{Barvainis}(1987)}]{Barvainis1987}
{Barvainis}, R. 1987, \apj, 320, 537, \dodoi{10.1086/165571}

\bibitem[{{Begelman} {et~al.}(1980){Begelman}, {Blandford}, \&
  {Rees}}]{Begelman1980}
{Begelman}, M.~C., {Blandford}, R.~D., \& {Rees}, M.~J. 1980, \nat, 287, 307,
  \dodoi{10.1038/287307a0}

\bibitem[{{Bentz} \& {Katz}(2015)}]{Bentz2015}
{Bentz}, M.~C., \& {Katz}, S. 2015, \pasp, 127, 67, \dodoi{10.1086/679601}

\bibitem[{{Burke-Spolaor} {et~al.}(2019){Burke-Spolaor}, {Taylor}, {Charisi},
  {Dolch}, {Hazboun}, {Holgado}, {Kelley}, {Lazio}, {Madison}, {McMann},
  {Mingarelli}, {Rasskazov}, {Siemens}, {Simon}, \&
  {Smith}}]{Burke-Spolaor2019}
{Burke-Spolaor}, S., {Taylor}, S.~R., {Charisi}, M., {et~al.} 2019, \aapr, 27,
  5, \dodoi{10.1007/s00159-019-0115-7}

\bibitem[{{Cappellari}(2013)}]{Cappellari2013}
{Cappellari}, M. 2013, \apjl, 778, L2, \dodoi{10.1088/2041-8205/778/1/L2}

\bibitem[{{Charisi} {et~al.}(2016){Charisi}, {Bartos}, {Haiman},
  {Price-Whelan}, {Graham}, {Bellm}, {Laher}, \& {M{\'a}rka}}]{Charisi2016}
{Charisi}, M., {Bartos}, I., {Haiman}, Z., {et~al.} 2016, \mnras, 463, 2145,
  \dodoi{10.1093/mnras/stw1838}

\bibitem[{{Colpi}(2014)}]{Colpi2014}
{Colpi}, M. 2014, \ssr, 183, 189, \dodoi{10.1007/s11214-014-0067-1}

\bibitem[{{Crook} {et~al.}(2007){Crook}, {Huchra}, {Martimbeau}, {Masters},
  {Jarrett}, \& {Macri}}]{Crook2007}
{Crook}, A.~C., {Huchra}, J.~P., {Martimbeau}, N., {et~al.} 2007, \apj, 655,
  790, \dodoi{10.1086/510201}

\bibitem[{{Davis} {et~al.}(2013){Davis}, {Bureau}, {Cappellari}, {Sarzi}, \&
  {Blitz}}]{Davis2013}
{Davis}, T.~A., {Bureau}, M., {Cappellari}, M., {Sarzi}, M., \& {Blitz}, L.
  2013, \nat, 494, 328, \dodoi{10.1038/nature11819}

\bibitem[{{De Rosa} {et~al.}(2019){De Rosa}, {Vignali}, {Bogdanovi{\'c}},
  {Capelo}, {Charisi}, {Dotti}, {Husemann}, {Lusso}, {Mayer}, {Paragi},
  {Runnoe}, {Sesana}, {Steinborn}, {Bianchi}, {Colpi}, {del Valle}, {Frey},
  {Gab{\'a}nyi}, {Giustini}, {Guainazzi}, {Haiman}, {Herrera Ruiz},
  {Herrero-Illana}, {Iwasawa}, {Komossa}, {Lena}, {Loiseau}, {Perez-Torres},
  {Piconcelli}, \& {Volonteri}}]{DeRosa2019}
{De Rosa}, A., {Vignali}, C., {Bogdanovi{\'c}}, T., {et~al.} 2019, \nar, 86,
  101525, \dodoi{10.1016/j.newar.2020.101525}

\bibitem[{{Desvignes} {et~al.}(2016){Desvignes}, {Caballero}, {Lentati},
  {Verbiest}, {Champion}, {Stappers}, {Janssen}, {Lazarus}, {Os{\l}owski},
  {Babak}, {Bassa}, {Brem}, {Burgay}, {Cognard}, {Gair}, {Graikou},
  {Guillemot}, {Hessels}, {Jessner}, {Jordan}, {Karuppusamy}, {Kramer},
  {Lassus}, {Lazaridis}, {Lee}, {Liu}, {Lyne}, {McKee}, {Mingarelli},
  {Perrodin}, {Petiteau}, {Possenti}, {Purver}, {Rosado}, {Sanidas}, {Sesana},
  {Shaifullah}, {Smits}, {Taylor}, {Theureau}, {Tiburzi}, {van Haasteren}, \&
  {Vecchio}}]{Desvignes2016}
{Desvignes}, G., {Caballero}, R.~N., {Lentati}, L., {et~al.} 2016, \mnras, 458,
  3341, \dodoi{10.1093/mnras/stw483}

\bibitem[{{Detweiler}(1979)}]{Detweiler1979}
{Detweiler}, S. 1979, \apj, 234, 1100, \dodoi{10.1086/157593}

\bibitem[{{D'Orazio} \& {Loeb}(2018)}]{2018ApJ...863..185D}
{D'Orazio}, D.~J., \& {Loeb}, A. 2018, \apj, 863, 185,
  \dodoi{10.3847/1538-4357/aad413}

\bibitem[{Ellis \& van Haasteren(2017)}]{evh17b}
Ellis, J., \& van Haasteren, R. 2017, jellis18/PTMCMCSampler: Official Release,
  \dodoi{10.5281/zenodo.1037579}

\bibitem[{{Ellis} {et~al.}(2017){Ellis}, {Vallisneri}, {Taylor}, \&
  {Baker}}]{enterprise}
{Ellis}, J.~A., {Vallisneri}, M., {Taylor}, S.~R., \& {Baker}, P.~T. 2017,
  https://github.com/nanograv/enterprise: enterprise.
\newblock \url{https://github.com/nanograv/enterprise}

\bibitem[{{Farris} {et~al.}(2014){Farris}, {Duffell}, {MacFadyen}, \&
  {Haiman}}]{Farris2014}
{Farris}, B.~D., {Duffell}, P., {MacFadyen}, A.~I., \& {Haiman}, Z. 2014, \apj,
  783, 134, \dodoi{10.1088/0004-637X/783/2/134}

\bibitem[{{Finn} \& {Thorne}(2000)}]{Finn2000}
{Finn}, L.~S., \& {Thorne}, K.~S. 2000, \prd, 62, 124021,
  \dodoi{10.1103/PhysRevD.62.124021}

\bibitem[{{Fitzpatrick}(1999)}]{Fitzpatrick1999}
{Fitzpatrick}, E.~L. 1999, \pasp, 111, 63, \dodoi{10.1086/316293}

\bibitem[{{Graham} {et~al.}(2015){Graham}, {Djorgovski}, {Stern}, {Drake},
  {Mahabal}, {Donalek}, {Glikman}, {Larson}, \& {Christensen}}]{Graham2015}
{Graham}, M.~J., {Djorgovski}, S.~G., {Stern}, D., {et~al.} 2015, \mnras, 453,
  1562, \dodoi{10.1093/mnras/stv1726}

\bibitem[{{Haehnelt} \& {Kauffmann}(2002)}]{Haehnelt2002}
{Haehnelt}, M.~G., \& {Kauffmann}, G. 2002, \mnras, 336, L61,
  \dodoi{10.1046/j.1365-8711.2002.06056.x}

\bibitem[{{Hellings} \& {Downs}(1983)}]{Hellings1983}
{Hellings}, R.~W., \& {Downs}, G.~S. 1983, \apjl, 265, L39,
  \dodoi{10.1086/183954}

\bibitem[{{Holgado} {et~al.}(2018){Holgado}, {Sesana}, {Sandrinelli}, {Covino},
  {Treves}, {Liu}, \& {Ricker}}]{Holgado2018}
{Holgado}, A.~M., {Sesana}, A., {Sandrinelli}, A., {et~al.} 2018, \mnras, 481,
  L74, \dodoi{10.1093/mnrasl/sly158}

\bibitem[{{Huchra} {et~al.}(2012){Huchra}, {Macri}, {Masters}, {Jarrett},
  {Berlind}, {Calkins}, {Crook}, {Cutri}, {Erdo{\v{g}}du}, {Falco}, {George},
  {Hutcheson}, {Lahav}, {Mader}, {Mink}, {Martimbeau}, {Schneider},
  {Skrutskie}, {Tokarz}, \& {Westover}}]{Huchra2012}
{Huchra}, J.~P., {Macri}, L.~M., {Masters}, K.~L., {et~al.} 2012, \apjs, 199,
  26, \dodoi{10.1088/0067-0049/199/2/26}

\bibitem[{{Jenet} {et~al.}(2004){Jenet}, {Lommen}, {Larson}, \&
  {Wen}}]{Jenet2004}
{Jenet}, F.~A., {Lommen}, A., {Larson}, S.~L., \& {Wen}, L. 2004, \apj, 606,
  799, \dodoi{10.1086/383020}

\bibitem[{{Kelley} {et~al.}(2017){Kelley}, {Blecha}, \&
  {Hernquist}}]{Kelley2017}
{Kelley}, L.~Z., {Blecha}, L., \& {Hernquist}, L. 2017, \mnras, 464, 3131,
  \dodoi{10.1093/mnras/stw2452}

\bibitem[{{Kelley} {et~al.}(2018){Kelley}, {Blecha}, {Hernquist}, {Sesana}, \&
  {Taylor}}]{Kelley2018}
{Kelley}, L.~Z., {Blecha}, L., {Hernquist}, L., {Sesana}, A., \& {Taylor},
  S.~R. 2018, \mnras, 477, 964, \dodoi{10.1093/mnras/sty689}

\bibitem[{{Kerr} {et~al.}(2020){Kerr}, {Reardon}, {Hobbs}, {Shannon},
  {Manchester}, {Dai}, {Russell}, {Zhang}, {van Straten}, {Os{\l}owski},
  {Parthasarathy}, {Spiewak}, {Bailes}, {Bhat}, {Cameron}, {Coles}, {Dempsey},
  {Deng}, {Goncharov}, {Kaczmarek}, {Keith}, {Lasky}, {Lower}, {Preisig},
  {Sarkissian}, {Toomey}, {Wang}, {Wang}, {Zhang}, \& {Zhu}}]{krh+20}
{Kerr}, M., {Reardon}, D.~J., {Hobbs}, G., {et~al.} 2020, arXiv e-prints,
  arXiv:2003.09780.
\newblock \doarXiv{2003.09780}

\bibitem[{{Kollmeier} {et~al.}(2017){Kollmeier}, {Zasowski}, {Rix}, {Johns},
  {Anderson}, {Drory}, {Johnson}, {Pogge}, {Bird}, {Blanc}, {Brownstein},
  {Crane}, {De Lee}, {Klaene}, {Kreckel}, {MacDonald}, {Merloni}, {Ness},
  {O'Brien}, {Sanchez-Gallego}, {Sayres}, {Shen}, {Thakar}, {Tkachenko},
  {Aerts}, {Blanton}, {Eisenstein}, {Holtzman}, {Maoz}, {Nandra}, {Rockosi},
  {Weinberg}, {Bovy}, {Casey}, {Chaname}, {Clerc}, {Conroy}, {Eracleous},
  {G{\"a}nsicke}, {Hekker}, {Horne}, {Kauffmann}, {McQuinn}, {Pellegrini},
  {Schinnerer}, {Schlafly}, {Schwope}, {Seibert}, {Teske}, \& {van
  Saders}}]{Kollmeier2017}
{Kollmeier}, J.~A., {Zasowski}, G., {Rix}, H.-W., {et~al.} 2017, arXiv
  e-prints, arXiv:1711.03234.
\newblock \doarXiv{1711.03234}

\bibitem[{{Kormendy} \& {Ho}(2013)}]{Kormendy2013}
{Kormendy}, J., \& {Ho}, L.~C. 2013, \araa, 51, 511,
  \dodoi{10.1146/annurev-astro-082708-101811}

\bibitem[{{Ma} {et~al.}(2014){Ma}, {Greene}, {McConnell}, {Janish},
  {Blakeslee}, {Thomas}, \& {Murphy}}]{Ma2014}
{Ma}, C.-P., {Greene}, J.~E., {McConnell}, N., {et~al.} 2014, \apj, 795, 158,
  \dodoi{10.1088/0004-637X/795/2/158}

\bibitem[{{McConnell} \& {Ma}(2013)}]{McConnell2013}
{McConnell}, N.~J., \& {Ma}, C.-P. 2013, \apj, 764, 184,
  \dodoi{10.1088/0004-637X/764/2/184}

\bibitem[{{Mingarelli} {et~al.}(2017){Mingarelli}, {Lazio}, {Sesana}, {Greene},
  {Ellis}, {Ma}, {Croft}, {Burke-Spolaor}, \& {Taylor}}]{Mingarelli2017}
{Mingarelli}, C.~M.~F., {Lazio}, T.~J.~W., {Sesana}, A., {et~al.} 2017, Nature
  Astronomy, 1, 886, \dodoi{10.1038/s41550-017-0299-6}

\bibitem[{{Mould} {et~al.}(2000){Mould}, {Huchra}, {Freedman}, {Kennicutt},
  {Ferrarese}, {Ford}, {Gibson}, {Graham}, {Hughes}, {Illingworth}, {Kelson},
  {Macri}, {Madore}, {Sakai}, {Sebo}, {Silbermann}, \& {Stetson}}]{Mould2000}
{Mould}, J.~R., {Huchra}, J.~P., {Freedman}, W.~L., {et~al.} 2000, \apj, 529,
  786, \dodoi{10.1086/308304}

\bibitem[{{Perera} {et~al.}(2019){Perera}, {DeCesar}, {Demorest}, {Kerr},
  {Lentati}, {Nice}, {Os{\l}owski}, {Ransom}, {Keith}, {Arzoumanian}, {Bailes},
  {Baker}, {Bassa}, {Bhat}, {Brazier}, {Burgay}, {Burke-Spolaor}, {Caballero},
  {Champion}, {Chatterjee}, {Chen}, {Cognard}, {Cordes}, {Crowter}, {Dai},
  {Desvignes}, {Dolch}, {Ferdman}, {Ferrara}, {Fonseca}, {Goldstein},
  {Graikou}, {Guillemot}, {Hazboun}, {Hobbs}, {Hu}, {Islo}, {Janssen},
  {Karuppusamy}, {Kramer}, {Lam}, {Lee}, {Liu}, {Luo}, {Lyne}, {Manchester},
  {McKee}, {McLaughlin}, {Mingarelli}, {Parthasarathy}, {Pennucci}, {Perrodin},
  {Possenti}, {Reardon}, {Russell}, {Sanidas}, {Sesana}, {Shaifullah},
  {Shannon}, {Siemens}, {Simon}, {Spiewak}, {Stairs}, {Stappers}, {Swiggum},
  {Taylor}, {Theureau}, {Tiburzi}, {Vallisneri}, {Vecchio}, {Wang}, {Zhang},
  {Zhang}, {Zhu}, \& {Zhu}}]{pdd+19}
{Perera}, B.~B.~P., {DeCesar}, M.~E., {Demorest}, P.~B., {et~al.} 2019, \mnras,
  490, 4666, \dodoi{10.1093/mnras/stz2857}

\bibitem[{{Ransom} {et~al.}(2019){Ransom}, {Brazier}, {Chatterjee}, {Cohen},
  {Cordes}, {DeCesar}, {Demorest}, {Hazboun}, {Lam}, {Lynch}, {McLaughlin},
  {Ransom}, {Siemens}, {Taylor}, \& {Vigeland}}]{2019BAAS...51g.195R}
{Ransom}, S., {Brazier}, A., {Chatterjee}, S., {et~al.} 2019, in Bulletin of
  the American Astronomical Society, Vol.~51, 195.
\newblock \doarXiv{1908.05356}

\bibitem[{{Rosado} {et~al.}(2015){Rosado}, {Sesana}, \& {Gair}}]{Rosado2015}
{Rosado}, P.~A., {Sesana}, A., \& {Gair}, J. 2015, \mnras, 451, 2417,
  \dodoi{10.1093/mnras/stv1098}

\bibitem[{{Sandrinelli} {et~al.}(2018){Sandrinelli}, {Covino}, {Treves},
  {Holgado}, {Sesana}, {Lindfors}, \& {Ramazani}}]{Sandrinelli2018}
{Sandrinelli}, A., {Covino}, S., {Treves}, A., {et~al.} 2018, \aap, 615, A118,
  \dodoi{10.1051/0004-6361/201732550}

\bibitem[{{Schutz} \& {Ma}(2016)}]{Schutz2016}
{Schutz}, K., \& {Ma}, C.-P. 2016, \mnras, 459, 1737,
  \dodoi{10.1093/mnras/stw768}

\bibitem[{{Sesana} {et~al.}(2018){Sesana}, {Haiman}, {Kocsis}, \&
  {Kelley}}]{Sesana2018}
{Sesana}, A., {Haiman}, Z., {Kocsis}, B., \& {Kelley}, L.~Z. 2018, \apj, 856,
  42, \dodoi{10.3847/1538-4357/aaad0f}

\bibitem[{{Seth} {et~al.}(2014){Seth}, {van den Bosch}, {Mieske}, {Baumgardt},
  {Brok}, {Strader}, {Neumayer}, {Chilingarian}, {Hilker}, {McDermid},
  {Spitler}, {Brodie}, {Frank}, \& {Walsh}}]{Seth2014}
{Seth}, A.~C., {van den Bosch}, R., {Mieske}, S., {et~al.} 2014, \nat, 513,
  398, \dodoi{10.1038/nature13762}

\bibitem[{{Shankar} {et~al.}(2016){Shankar}, {Bernardi}, {Sheth}, {Ferrarese},
  {Graham}, {Savorgnan}, {Allevato}, {Marconi}, {L{\"a}sker}, \&
  {Lapi}}]{Shankar_2016}
{Shankar}, F., {Bernardi}, M., {Sheth}, R.~K., {et~al.} 2016, \mnras, 460,
  3119, \dodoi{10.1093/mnras/stw678}

\bibitem[{{Siwek} {et~al.}(2020){Siwek}, {Kelley}, \& {Hernquist}}]{Siwek2020}
{Siwek}, M.~S., {Kelley}, L.~Z., \& {Hernquist}, L. 2020, arXiv e-prints,
  arXiv:2005.09010.
\newblock \doarXiv{2005.09010}

\bibitem[{{Sudou} {et~al.}(2003){Sudou}, {Iguchi}, {Murata}, \&
  {Taniguchi}}]{Sudou2003}
{Sudou}, H., {Iguchi}, S., {Murata}, Y., \& {Taniguchi}, Y. 2003, Science, 300,
  1263, \dodoi{10.1126/science.1082817}

\bibitem[{{Taylor} {et~al.}(2019){Taylor}, {Burke-Spolaor}, {Baker}, {Charisi},
  {Islo}, {Kelley}, {Madison}, {Simon}, {Vigeland}, \& {Nanograv
  Collaboration}}]{2019BAAS...51c.336T}
{Taylor}, S., {Burke-Spolaor}, S., {Baker}, P.~T., {et~al.} 2019, \baas, 51,
  336.
\newblock \doarXiv{1903.08183}

\bibitem[{{Taylor} {et~al.}(2016){Taylor}, {Vallisneri}, {Ellis}, {Mingarelli},
  {Lazio}, \& {van Haasteren}}]{Taylor2016}
{Taylor}, S.~R., {Vallisneri}, M., {Ellis}, J.~A., {et~al.} 2016, \apjl, 819,
  L6, \dodoi{10.3847/2041-8205/819/1/L6}

\bibitem[{{Thomas} {et~al.}(2016){Thomas}, {Ma}, {McConnell}, {Greene},
  {Blakeslee}, \& {Janish}}]{Thomas2016}
{Thomas}, J., {Ma}, C.-P., {McConnell}, N.~J., {et~al.} 2016, \nat, 532, 340,
  \dodoi{10.1038/nature17197}

\bibitem[{{Tully} {et~al.}(2016){Tully}, {Courtois}, \& {Sorce}}]{Tully2016}
{Tully}, R.~B., {Courtois}, H.~M., \& {Sorce}, J.~G. 2016, \aj, 152, 50,
  \dodoi{10.3847/0004-6256/152/2/50}

\bibitem[{{Tully} {et~al.}(2009){Tully}, {Rizzi}, {Shaya}, {Courtois},
  {Makarov}, \& {Jacobs}}]{Tully2009}
{Tully}, R.~B., {Rizzi}, L., {Shaya}, E.~J., {et~al.} 2009, \aj, 138, 323,
  \dodoi{10.1088/0004-6256/138/2/323}

\bibitem[{{Volonteri} {et~al.}(2003){Volonteri}, {Haardt}, \&
  {Madau}}]{Volonteri2003}
{Volonteri}, M., {Haardt}, F., \& {Madau}, P. 2003, \apj, 582, 559,
  \dodoi{10.1086/344675}

\bibitem[{{Walsh} {et~al.}(2015){Walsh}, {van den Bosch}, {Gebhardt},
  {Yildirim}, {G{\"u}ltekin}, {Husemann}, \& {Richstone}}]{Walsh2015}
{Walsh}, J.~L., {van den Bosch}, R. C.~E., {Gebhardt}, K., {et~al.} 2015, \apj,
  808, 183, \dodoi{10.1088/0004-637X/808/2/183}

\bibitem[{{Walsh} {et~al.}(2016){Walsh}, {van den Bosch}, {Gebhardt},
  {Y{\i}ld{\i}r{\i}m}, {Richstone}, {G{\"u}ltekin}, \& {Husemann}}]{Walsh2016}
---. 2016, \apj, 817, 2, \dodoi{10.3847/0004-637X/817/1/2}

\bibitem[{{Weinzirl} {et~al.}(2009){Weinzirl}, {Jogee}, {Khochfar}, {Burkert},
  \& {Kormendy}}]{Weinzirl2009}
{Weinzirl}, T., {Jogee}, S., {Khochfar}, S., {Burkert}, A., \& {Kormendy}, J.
  2009, \apj, 696, 411, \dodoi{10.1088/0004-637X/696/1/411}

\bibitem[{{Zhu} {et~al.}(2014){Zhu}, {Hobbs}, {Wen}, {Coles}, {Wang},
  {Shannon}, {Manchester}, {Bailes}, {Bhat}, {Burke-Spolaor}, {Dai}, {Keith},
  {Kerr}, {Levin}, {Madison}, {Os{\l}owski}, {Ravi}, {Toomey}, \& {van
  Straten}}]{Zhu2014}
{Zhu}, X.-J., {Hobbs}, G., {Wen}, L., {et~al.} 2014, \mnras, 444, 3709,
  \dodoi{10.1093/mnras/stu1717}

\end{thebibliography}
\bibliographystyle{aasjournal}



\end{document}